\documentclass[10pt]{article}
\usepackage{amsmath}
\usepackage{graphicx}
\usepackage{amsfonts}
\usepackage{amssymb}
\usepackage{epsf}
\usepackage{latexsym}

\usepackage{fix-cm}

\def\80{\hspace{0.8in}}

\newcommand{\be}{\begin{enumerate}}
\newcommand{\ee}{\end{enumerate}}
\newcommand{\bi}{\begin{itemize}}
\newcommand{\ei}{\end{itemize}}
\newcommand{\bd}{\begin{description}}
\newcommand{\ed}{\end{description}}

\def\beq{\begin{equation}}
\def\eeq{\end{equation}}
\def\bea{\begin{eqnarray}}
\def\eea{\end{eqnarray}}
\def\foo{\footnote}
\def\hat{\widehat}
\def\tilde{\widetilde}

\def\pa{\partial}
\def\d{\textrm{d}}

\def\ttA{\mbox{\tt A}}
\def\ttB{\mbox{\tt B}}
\def\ttC{\mbox{\tt C}}
\def\ttH{\mbox{\tt H}}
\def\ttL{\mbox{\tt L}}
\def\ttD{\mbox{\tt D}}

\def\cr{\mbox{\scriptsize{\bf $\mbox{ } \times \mbox{ }$}}}

\def\mI{\mbox{I}}
\def\mJ{\mbox{J}}

\def\mL{\mbox{L}}

\def\mN{\mbox{N}} 

\def\mP{\mbox{P}}
\def\mQ{\mbox{Q}}
\def\mR{\mbox{R}}

\def\sa{\mbox{\scriptsize a}}
\def\sb{\mbox{\scriptsize b}}

\def\sd{\mbox{\scriptsize d}}
\def\se{\mbox{\scriptsize e}}
\def\sf{\mbox{\scriptsize f}}
\def\sg{\mbox{\scriptsize g}} 
 
\def\si{\mbox{\scriptsize i}}

\def\sll{\mbox{\scriptsize l}}  
\def\sm{\mbox{\scriptsize m}}
\def\sn{\mbox{\scriptsize n}} 
\def\so{\mbox{\scriptsize o}} 
\def\sp{\mbox{\scriptsize p}}
\def\sq{\mbox{\scriptsize q}}
\def\sr{\mbox{\scriptsize r}}
\def\sss{\mbox{\scriptsize s}}
\def\st{\mbox{\scriptsize t}}

\def\sF{\mbox{\scriptsize F}}
\def\sG{\mbox{\scriptsize G}}

\def\sI{\mbox{\scriptsize I}}

\def\sN{\mbox{\scriptsize N}} 
\def\s-{\mbox{\scriptsize -}}

\def\sR{\mbox{\scriptsize R}}
\def\sS{\mbox{\scriptsize S}}

\def\barp{\bar{p}}
\def\barq{\bar{q}}
\def\barr{\bar{r}}
\def\eph(B){\mbox{\scriptsize emergent(LMB)}}

\def\tn{\mbox{\tiny n}}

\def\tJ{\mbox{\tiny J}}

\def\tN{\mbox{\tiny N}}

\def\fA{\mbox{\sffamily A}}

\def\fE{\mbox{\sffamily E}}
\def\fF{\mbox{\sffamily F}}

\def\fH{\mbox{\sffamily H}}

\def\fP{\mbox{\sffamily P}}
\def\fQ{\mbox{\sffamily Q}}
\def\fR{\mbox{\sffamily R}}
\def\fS{\mbox{\sffamily S}}
\def\fT{\mbox{\sffamily T}}
\def\fU{\mbox{\sffamily U}}
\def\fV{\mbox{\sffamily V}}

\def\sfA{\mbox{\sffamily{\scriptsize A}}}
\def\sfB{\mbox{\sffamily{\scriptsize B}}}
\def\sfC{\mbox{\sffamily{\scriptsize C}}}

\def\sfE{\mbox{\sffamily{\scriptsize E}}}

\def\sfT{\mbox{\sffamily{\scriptsize T}}}
\def\sfU{\mbox{\sffamily{\scriptsize U}}}

\def\q{\underline{q}}

\def\tip{\tilde{p}}
\def\tiq{\tilde{q}}

\begin{document}
\begin{titlepage}
\vspace{.7in}
\begin{center}

\LARGE{\bf TRIANGLELAND. I. CLASSICAL DYNAMICS WITH} 

\vspace{0.1in}

{\bf EXCHANGE OF RELATIVE ANGULAR MOMENTUM}\normalsize

\vspace{.4in}

\large{\bf Edward Anderson}$^{1}$

\vspace{.2in}

\large{\em Peterhouse, Cambridge CB2 1RD } and \normalsize 

\vspace{.2in}

\large{\em DAMTP, Centre for Mathematical Sciences, Wilberforce Road, Cambridge CB3 OWA.}

\end{center}

\begin{abstract}

In Euclidean relational particle mechanics, only relative times, relative angles and relative separations 
are meaningful.  
Barbour--Bertotti (1982) theory is of this form and can be viewed as a recovery of (a portion of) 
Newtonian mechanics from relational premises.  
This is of interest in the absolute versus relative motion debate and also shares a number 
of features with the geometrodynamical formulation of general relativity, making it suitable for some 
modelling of the problem of time in quantum gravity.  
I also study similarity relational particle mechanics (`dynamics of pure shape'), 
in which only relative times, relative angles and {\sl ratios of} relative separations are meaningful.  
This I consider firstly as it is simpler, particularly in 1 and 2 d, for which the 
configuration space geometry turns out to be well-known, e.g. $\mathbb{S}^2$ for the `triangleland' 
(3-particle) case that I consider in detail.
Secondly, the similarity model occurs as a sub-model within the Euclidean model: that admits a shape--scale 
split.

For harmonic oscillator like potentials, similarity triangleland model turns out to have the same 
mathematics as a family of rigid rotor problems, while the Euclidean case turns out to have parallels 
with the Kepler--Coulomb problem in spherical and parabolic coordinates.      
Previous work on relational mechanics covered cases where the constituent subsystems do not exchange 
relative angular momentum, which is a simplifying (but in some ways undesirable) feature paralleling 
centrality in ordinary mechanics. 
In this paper I lift this restriction.
In each case I reduce the relational problem to a standard one, thus obtain various exact, asymptotic 
and numerical solutions, and then recast these into the original mechanical variables for physical 
interpretation.  

 
\end{abstract}

\vspace{1in}


\mbox{ }

\vspace{3in}

\noindent$^1$ ea212@cam.ac.uk


\end{titlepage}

\section{Introduction}

The traditional formulation of mechanics \cite{Principia} has rested on absolute space and absolute time.  
However, Leibniz \cite{LCC} and Mach \cite{Mach} raised philosophically well-motivated relational 
objections to these foundations. 
(See e.g. \cite{Berkeley, B86, buckets, Comments} for further discussion of this `absolute 
versus relative motion debate'.)  
It is reasonable to consider whether relational principles apply to physics as a whole.  
However, for many years no means were known by which physical theories could be built along these lines. 
Then Barbour and Bertotti \cite{BB82} and Barbour \cite{B03} found some relational particle mechanics 
[Reissner's earlier theory -- subsequently rediscovered by Schr\"{o}dinger and by Barbour--Bertotti --  
\cite{BB77, buckets} is incompatible with mass-anisotropy experiments.]

Note that the present paper uses the word `relational' in Barbour's sense.  
This is worth some discussion because Rovelli \cite{Rovelli} uses the same word, but each of he and 
Barbour take it to mean something different.  
In outline, Rovelli's classical relationalism involves objects not being located in spacetime but being 
located with respect to each other. 
Rovelli also has a quantum relationalism that quantum states for a particular subsystem only make 
sense with respect to another subsystem. 
He then speculates that these two relationalisms of his might be related (p 157 of the online version 
of \cite{Rovelli}: ``Is there a connection... This is of course very vague, and might lead
nowhere, but I find the idea intriguing." 
Barbour on the other hand, has specific spatial and temporal relationalism postulates that embody 
particular ideas of Mach (that time is to be abstracted from change) and Leibniz (the identity of 
indiscernibles), each of which is sharply implemented by particular mathematics at the classical level, 
as follows.   

\noindent
A physical theory is {\it temporally relational} if there is no meaningful primary notion of time for 
the whole system thereby described (e.g. the universe) \cite{BB82, RWR, FORD}.  
This is implemented by using actions that are manifestly reparametrization invariant while also being 
free of extraneous time-related variables [such as Newtonian time or General Relativity (GR)'s lapse].   
This reparametrization invariance then directly produces primary constraints quadratic in the momenta 
(such as the energy constraint of mechanics or the Hamiltonian constraint of GR).  

\noindent
A physical theory is {\it configurationally relational} if a certain group $G$ of transformations that 
act on the theory's configuration space $\fQ$ are physically meaningless \cite{BB82, RWR, Lan, FORD}.  
This can be is implemented by such as\footnote{This 
is my own \cite{Lan, Phan, Lan2} passive, mathematician's implementation, whereas Barbour thinks about 
this in active terms that physicists sometimes use.  
This difference in thinking does not, however, lead to any tangible discrepancies in the material 
of this paper.} 
using arbitrary-$G$-frame-corrected quantities rather than `bare' $\fQ$-configurations.
For, despite this augmenting $\fQ$ to the principal bundle $P(\fQ, G)$, variation with respect to each 
adjoined independent auxiliary $G$-variable produces a secondary constraint linear in the momenta 
(e.g. the GR momentum constraint arises as a vectorial collection of 3 such constraints) which removes 
one $G$ degree of freedom and one redundant degree of freedom among the $\fQ$ variables. 
Thus one ends up dealing with the desired reduced configuration space -- the quotient space $\fQ/G$.  
Configurational relationalism includes as subcases both spatial relationalism and internal relationalism 
(in the sense of gauge theory).  
Configurational relationalism can also be implemented, at least in some cases \cite{TriCl, FORD}, by 
working directly on reduced configuration space. (Reissner's theory was formulated along these direct 
lines, while I also found that the theories \cite{BB82, B03} can be arrived at in 1- and 2-d from a 
direct implementation \cite{FORD}.)

One difference between Barbour and Rovelli's approaches is as follows (another is mentioned in 
Subsec 2.1).
In Sec. 2.4.4 of \cite{Rovelli}, Rovelli discusses ``meanings of time" and identifies time in Newtonian 
physics as a metric line that exists alongside the configuration space of the system, nowhere 
reflecting Barbour's starting point that time is derived from change alongside consideration of
the configuration spaces alone, on which Jacobi-type variational principles are defined.  
This then takes Barbour straight to the situation in which no variable is distinguished as time at the 
kinematic level; while Rovelli characterizes this as an essential difference between non-relativistic 
and relativistic mechanics in his approach, in Barbour's approach this distinction has dissolved.  
The recovery of Newtonian dynamics from relational particle models also has features by which such 
models are closer in objective structure to GR than Rovelli generally holds nonrelativistic mechanics 
models to be in \cite{Rovelli}.  
Thereby, relational particle models provide tractable models outside the scheme in \cite{Rovelli}.

As far as I know, Barbour and collaborators have not as yet reached a specific notion of quantum 
relationalism.  
Rovelli's own idea of relationalism at the quantum level above-mentioned does share a number of features 
with record-theoretic positions that Barbour and I have previously advocated \cite{B94II, EOT, Records}.
It is not as yet know whether such records-theoretic positions really have any substantial conceptual 
or technical ties to Barbour's notion of classical relationalism, any more than Rovelli knows whether 
his similar quantum position to be tied to his own classical relationalism (as above-quoted).

While Barbour's relationalism has the virtue of specifically being in line with Leibniz and Mach 
(whereby alongside having attained a concrete mathematical implementation of these ideas it is
definitely of interest to the foundations of physics and theoretical physics and so deserves full 
investigation), I would not dismiss the possibility that Rovelli's relationalism is {\it also} in line 
with other (interpretations of) Leibniz, Mach or other such historical figures, and, in any case, 
has original value and is useful in a major quantum gravity program (Loop Quantum Gravity).

{\sl The current paper and its sequel \cite{08II} concern technical advances with examples specifically 
of Barbour's relational program} -- I subsequently use `relational' in this sense except where I 
specifically say otherwise.  
Sec 2 presents Euclidean relational particle mechanics \cite{BB82, B86, BS89, B94I, GGM, Paris, 06I, 
TriCl, FORD} (also referred to as {\it scaled} models) and similarity relational particle mechanics 
\cite{B03, Paris, 06II, TriCl, FORD} (also referred to as {\it scalefree} models) and further motivates 
the relational scheme by arguing that some (conformo)geometrodynamical formulations of GR can be 
regarded as arising therein too.
Sec 3 explains the configuration space structure of relational particle mechanics which has further parallels with GR.  
All these parallels are eventually relevant as regards relational particle mechanics furbishing useful toy models 
\cite{K92, B94II, EOT, 06I, 06II, SemiclI,  Records, New, BF08} for the Problem 
of Time in Quantum Gravity \cite{K92, I93} and other issues in the foundations of Quantum Cosmology 
\cite{EOT, Halliwell03}.    
Use of relational particle mechanics in both the absolute versus relative motion debate and the study of conceptual strategies 
suggested toward resolving the Problem of Time in Quantum Gravity would benefit from having a good 
working understanding of explicit examples of relational particle mechanics, 
which is the subject of this paper at the classical level.

Relational particle models concern N particles in dimension d.  
As the general configuration for d = 3 is an N-haedron, I term the 3-d N-particle relational particle 
model {\it N-haedronland}.  
Likewise, I term the 2-d N-particle relational particle model {\it N-a-gonland}, and the 1-d one 
{\it N-stop metroland} (as in urban public transport maps).  
The special 3-a-gonland considered as the principal example in this paper I term {\it triangleland}.

Noting that scalefree N-stop metroland and N-a-gonland have fairly standard geometry 
($\mathbb{S}^{\sN - 2}$ and $\mathbb{CP}^{\sN - 2}$ respectively \cite{FORD}) permits explicit 
reductions \cite{TriCl, FORD} and 
subsequent availability of useful coordinate systems and methods of mathematical physics.  
For scalefree triangleland, $\mathbb{CP}^{1}  = \mathbb{S}^{2}$, so one has `twice as many 
techniques', so in this paper I choose to study this case.  
Scaled triangleland also permits explicit reduction and its configuration space in shape-scale variables 
takes the form $C(\mathbb{S}^{2})$, which also makes for tractable and interesting explicit examples, 
where the {\it cone} \cite{Cones} C(X) over a space X is 
$\mathbb{R}_+ \times X \mbox{ } \bigcup \mbox{ } 0$, the special cone-point or `apex'.

In Sec 4 I give Euler--Lagrange equations for scalefree N-stop metroland and N-a-gonland and give 
further specific forms for the exceptional triangleland case.  
I then consider examples with potentials that are independent of the relative angle $\Phi$\footnote{The 
notation for this and the next paragraph is as follows.  
$\underline{\mR}_i$, $i$ = 1, 2, are {\it relative Jacobi coordinates} \cite{Marchal}.  
$\Phi$ is the angle between these. 
By $\underline{\iota}_i$ being mass-weighted, I mean that $\underline{\iota}_i = 
\sqrt{\mu_i}\underline{R}_i$, where $\mu_i$ are the particle (cluster) masses associated with 
$\underline{\mR}_i$ \cite{Marchal}.  
$\iota_i = ||\underline{\iota}_i||$, the magnitude of $\underline{\iota}_i$, and 
$I_i = \iota_i^2 = \mu_iR_i^2$, the ith Jacobi (barycentric) partial moment of inertia. 
${\cal R}$ is the simple ratio variable $\iota_1/\iota_2$ and $\Theta = 2\mbox{arctan}{\cal R}$, which 
turns out to geometrically be the azimuthal spherical angle; see also Fig 1 and Secs 2--3 for further 
interpretation and depictions of these.}
as well as harder ones that depend on this.   
(The former is a substantial simplification \cite{TriCl} -- in close analogy with centrality in ordinary 
mechanics -- but it is the harder latter case that is relevant to various Problem of Time schemes -- 
semiclassical emergent time and possibly the semblance of dynamics in timeless records schemes discussed 
in paper II \cite{08II}.)  
In particular, I consider the general case of harmonic oscillator like potentials between all particles, looking at the 
`special' ($\Phi$-independent) and `very special' (constant) subcases within as well as the general 
small and large asymptotic behaviour, and recast this in terms of the problem's `original variables' -- 
{\it mass-weighted relative Jacobi variables} 
$\underline{\iota}_1$, $\underline{\iota}_2$ that still contain an absolute orientation.  
Identifying the `very special' problem's mathematics as corresponding to the linear rigid rotor, 
the `special' problem's as that with additionally a background homogeneous electric field in the axial 
direction and the general problem's likewise but now with a general direction, is valuable in this 
paper and in subsequent work at the QM level in paper II.   
In Sec 5 I use that a rotation (or, equivalently, a normal modes construction) maps the general case to 
the special case for this particular problem.  
Thus I can get as far in solving for the general case as I can with the `special' case.

In Sec 6, I consider scaled triangleland's Euler--Lagrange equations in terms of the 
straightforwardly relational variables $(\iota_1, \iota_2, \Phi)$, the useful ($I_1, I_2, \Phi$) 
coordinates (that turn out to be parabolic coordinates), the $C(\mathbb{CP}^1)$ 
presentation shape--scale coordinates $(I, {\cal R}, \Phi)$ and the $C(\mathbb{S}^2)$ 
presentation shape--scale coordinates $(I, \Theta, \Phi)$.  
I then exactly solve the `special' case of multi-harmonic oscillator like potential for scaled triangleland, 
by mapping it in $(I_1, I_2, \Phi)$ coordinates to a close analogue of the Kepler--Coulomb problem  
(which move remains useful at the quantum level \cite{08III}).  
I then give Euclidean relational particle mechanics' own rotation/normal modes construction, whereby my having obtained the special 
solution enables me to also obtain the general solution, albeit this paper only has room for 
presenting the special solution and its physical interpretation.

I conclude in Sec 7, including an outline of further promising relational particle mechanics examples.   
Paper II considers this paper's similarity models at the quantum level and \cite{08III} 
likewise for this paper's Euclidean models.  
Interesting Problem of Time in Quantum Gravity applications of these models will be developed in yet 
further papers \cite{New}.

\section{Examples of relational theories}

\subsection{Euclidean relational particle mechanics} 

In {\it Euclidean}, or {\it scaled}, {\it relational particle mechanics} \cite{BB82, B86, B94I, EOT}, only 
relative times, relative angles and relative separations are meaningful.    
E.g. for 3 particles in dimension d $> 1$, Euclidean relational particle mechanics is a dynamics of the 
triangle that the 3 particles form.    
Euclidean relational particle mechanics was originally \cite{BB82} conceived for $\fQ = \fQ(\mN, \d) = 
\mathbb{R}^{\sN\sd}$ the positions $q_{I\alpha}$ of N particles in d-dimensional space with $G$ the 
d-dimensional Euclidean group Eucl(d) 
of translations, Tr(d), and rotations, Rot(d).\footnote{Lower-case Greek letters are spatial indices.
Capital indices are used for the N particle position coordinates, 
lower-case indices for n = N -- 1 coordinates describing relative particle (cluster) separations, 
barred and tilded lower-case indices to take values 1 to n -- 1 values and 
hatted lower-case indices to take values 1 to n -- 2.
The dot denotes ${\d}/{\d\lambda}$ for $\lambda$ a label-time parameter that has no physical 
meaning since (\ref{action}) is reparametrization-invariant.  
$B_{\alpha}$ generates the rotations.  
As well as the obvious 3-d case, the 1 and 2 d cases are incorporable into this form as follows.  
Take $B_{\alpha}$ = (0, 0, $B$) in 2-d.  
Then $\ttL^{\alpha}$ has just one component that is nontrivially zero.  
Take furthermore $B$ = 0 in 1-d.  
Then there is no $\ttL^{\alpha}$ constraint at all.  
$M^{ij\alpha\beta}$ is the kinetic metric $\mu_i\delta^{ij}\delta^{\alpha\beta}$, where $\mu_i$ are the 
Jacobi masses \cite{Marchal}. 
[E.g. for this paper's triangleland example, these are 
$\mu_1 = m_2m_3/\{m_2 + m_3\}$ and $\mu_2 = m_1\{m_2 + m_3\}/\{m_1 + m_2 + m_3\}$ for $m_I$ the 
particle masses.].  
$N_{ij\alpha\beta}$ is the inverse of this array.}
%
However, eliminating Tr(d) is trivial and produces a theory of essentially the same form as the original 
if relative Jacobi coordinates $R_{i\alpha}$ are employed \cite{06I}, so I take that as my starting 
position.  
Relative Jacobi coordinates are inter-particle (cluster) separations chosen such that the kinetic term is 
diagonal \cite{Marchal}; thus I take my $\fQ$ to be $\fR(\mN, \d)$, the space of relative separations 
(this is $\mathbb{R}^{\sn\sd}$), and $G$ to be Rot(d).  
Then configurational relationalism takes the form that one is to construct one's action using the 
arbitrary Rot(d) frame expressions $\circ R_{i\alpha} \equiv \dot{R}_{i\alpha} - 
{\epsilon_{\alpha}}^{\beta\gamma}\dot{B}_{\beta}R_{i\gamma}$ rather than `bare' $\dot{R}_{i\alpha}$.

\mbox{ }

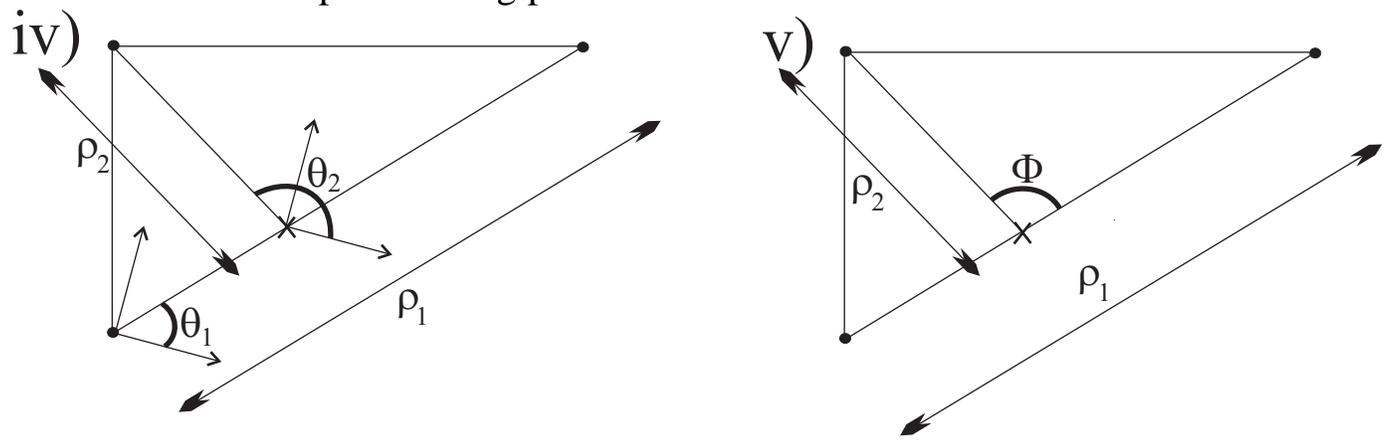
\noindent{\footnotesize [{\bf Figure 1}: Coordinate systems for scaled triangleland.   

\noindent i) Absolute particle position coordinates 
($\underline{q}_1$, $\underline{q}_2$, $\underline{q}_3$) with respect to fixed axes and a fixed origin O; the corresponding particle 
masses are $m_I$, $I$ = 1 to 3.

\noindent ii) Relative particle position coordinates, any 2 of which form a basis.     
 
\noindent iii) Relative Jacobi coordinates ($\underline{R}_1$, $\underline{R}_2$); the corresponding Jacobi masses are $\mu_i$, $i$ = 1, 2.   
The mass-scaled relative Jacobi coordinates are related to these by $\underline{\iota}_i = 
\sqrt{\mu_i}\underline{R}_i$.  

\noindent iv) Bipolar relative Jacobi coordinates ($\rho_1$, $\theta_1$, $\rho_2$, $\theta_2$).   
The mass-scaled radial Jacobi coordinates are $\iota_i = \sqrt{\mu}\rho_i$. 
These coordinates still refer to fixed axes.  

\noindent v) Fully relational coordinates ($\rho_1, \rho_2, \Phi$), for $\Phi$ = 
arccos$\left(\frac{\underline{R}_1\cdot\underline{R}_2}{||\underline{R}_1||||\underline{R}_2||}\right)$ 
the relational `Swiss army knife angle' between the 2 relative Jacobi vectors.   
The coordinate ranges are $0 \leq \rho_i < \infty$, $0 \leq \Phi < 2\pi$.  ]} 

\mbox{ }

The action that one builds is of Jacobi type \cite{Lanczos} so as to implement temporal relationalism,  
\beq
\fS^{}_{}[R_{i\alpha}, \dot{R}_{i\alpha}, \dot{B}_{\alpha}] = 
2\int\d\lambda\sqrt{\fT^{}\{\fU + \fE\}} \mbox{ } .     
\label{action}
\eeq
Here the kinetic term $\fT^{}(R_{i\alpha}, \dot{R}_{i\alpha}, \dot{B}_{\alpha}) = 
M^{ij\alpha\beta}\circ R_{i\alpha}\circ R_{j\beta}/2$ is homogeneous quadratic in the velocities. 
$\fU$ is minus the potential energy $\fV$ which is a function of 
$\sqrt{\underline{R}_i\cdot\underline{R}_j}$ alone, and $\fE$ is the total energy of the closed 
system/universe; hitherto this has been taken to have a fixed value (but should it: no observer would 
know it exactly?) 
[Note that each such action $\fS_{\tJ} = 2\int\d\lambda\sqrt{\fT\{\fE + \fU\}}$ is indeed equivalent to 
the more well known Euler--Lagrange actions
\beq
\fS = \int\d t\{\fT - \fV\} \mbox{ } ,
\label{Lagaction}
\eeq 
where $\fT$ now differs from the its previous form in containing $*$ in place of $\dot{\mbox{ }}$.  
See e.g. \cite{Lanczos} for obtaining actions of type (\ref{action}) from actions of type 
(\ref{Lagaction}) by parametrization and Routhian reduction to eliminate $\d t/\d\lambda$.]

From (\ref{action}), the conjugate momenta are then
\beq
P^{i\alpha} =  M^{ij\alpha\beta}*R_{j\beta} \mbox{ } \mbox{ for } \mbox{ } 
*R_{i\alpha} \equiv {R_{i\alpha}}^* - \epsilon_{\alpha}\mbox{}^{\beta\gamma}{B_{\beta}}^*R_{i\gamma} 
\label{Rmom}
\eeq
and $\mbox{}^* \equiv \sqrt{\{\fU + \fE\}/{\fT^{}}}\mbox{ }\dot{\mbox{}} = {\d}/{\d t}$ for $t$ the 
emergent `Leibniz--Mach--Barbour' time that coincides here with Newtonian time.  
This object introduced, another difference between Barbour's scheme and Rovelli's can be pointed out.  
In Rovelli's scheme, one clock/timestandard is much as good as another (at the conceptual level rather 
than what is convenient or accurate), while in Barbour's scheme there emerges a Leibniz--Mach--Barbour 
timestandard to which everything in the universe contributes (an `ephemeris' timestandard 
\cite{B94I,SemiclI,fqxi}). 
This choice of time is distinguished by its substantially simplifying both the above momentum-velocity 
relations and the Euler--Lagrange equations, and it amounts to an emergent recovery of other notions of 
time like Newtonian time here in mechanics, or proper time or cosmic time in Sec 2.3, and it has some
parallels with the actual ephemeris time in official use in the first half of the 20th century 
\cite{Clemence}, which is based on the totality of the motions of the objects in the solar system.

Reparametrization invariance implies \cite{Dirac} that these must obey at least one primary constraint;  
here there is one, which takes the form of an energy constraint 
\beq
\ttH \equiv N_{ij\alpha\beta}P^{i\alpha}P^{j\beta}/2 + \fV = \fE  
\label{En} \mbox{ } , 
\eeq
to which the momenta contribute quadratically but not linearly.  
Variation with respect to $B_{\alpha}$ yields as a secondary constraint the zero total angular 
momentum constraint  
\beq
{\ttL}^{\alpha} \equiv {\epsilon^{\alpha\beta}}_{\gamma}R_{i\beta}P^{i\gamma} = 0 \mbox{ } ,  
\label{ZAM}
\eeq
which is linear in the momenta and is interpretable as the physical content of the theory being in 
relative separations and relative angles and not in absolute angles.

Euclidean relational particle mechanics is Leibnizian/Machian in form, and yet is in agreement with a subset of Newtonian mechanics -- the 
zero total angular momentum universes.
The recovery from relationalism of many of the results previously obtained by standard absolutist 
physics and the finding of similar mathematical structures along both routes (at least in the simpler 
cases) is an interesting reconciliation as regards the extent to which we have not been prejudiced by 
studying absolutist physics. 
This paper and \cite{08II} show that this trend extends further, albeit the mathematics arising from 
similarity relational particle mechanics is distinct from that in the ordinary central force problem in 
having restricted, unusual potentials inherited from the scale invariance).  
Sec 2.3 and 2.4 recollect that a similar trend is also present in GR.

\subsection{Similarity relational particle mechanics}

In {\it similarity}, or {\it scalefree}, {\it relational particle mechanics} \cite{B03, 06II, TriCl}, only relative times, relative angles and ratios of relative separations 
are meaningful.    
I.e., it is a dynamics of shape excluding size: a {\sl dynamics of pure shape}.  
In the case of 3 particles in dimension d $> 1$, similarity relational particle mechanics is the dynamics of the shape of the 
triangle that the 3 particles form.      
$G$ is now augmented by the dilations Dil(d) = Dil to form the Similarity group Sim(d), so that 
$\circ R_{i\alpha}$ now takes the form $\circ R_{i\alpha} \equiv \dot{R}_{i\alpha} - 
{\epsilon_{\alpha}}^{\beta\gamma}\dot{B}_{\beta}R_{i\gamma} + \dot{C}R_{i\alpha}$, where $C$ generates 
the dilations.  
Noting that the `banal conformal transformation',\footnote{Performing this 
transformation  clearly leaves invariant product-type actions and so of Jacobi-type actions 
(\ref{action}) and its similarity relational particle mechanics counterpart and of the reduced actions that follow from these in Sec 3, and 
of their GR counterparts such as the Baierlein--Sharp--Wheeler action \cite{BSW}, (\ref{GRaction}), and (\ref{GRconfaction}) and its 
variants.
$\fT$ and $\fE + \fU$ scale compensatingly and then deduce from $\mbox{}^* \equiv \sqrt{\{\sfE + \sfU\}/{\sfT}}
\mbox{ }\dot{\mbox{}}$ that $*$ scales as $* \longrightarrow *_{\Omega} = \Omega^{-2}*$.
(Euler--Lagrange or Arnowitt--Deser--Misner type actions have this invariance too but its presence therein 
is less obvious to spot \cite{Banal}.)
I term each particular choice of $\Omega$ a `banal conformal representation'.  
Note: classically Euler--Lagrange equations are invariant \cite{Banal} so applying a banal conformal transformation makes no 
difference and it is just a means of computational convenience in some cases.  
But retaining this lack of difference at the quantum level has implications \cite{Banal, 08II, 08III}.}  
\beq
\fT \rightarrow \fT_{\Omega} = \Omega^2\fT \mbox{ } \mbox{ } , \mbox{ } \mbox{ } 
\fE + \fU \rightarrow \fE_{\Omega} + \fU_{\Omega} = \{\fE - \fV\}/\Omega^2 \mbox{ } , 
\eeq 
leaves the Jacobi action invariant, the most natural presentation \cite{TriCl} for the action is 
\beq
\fS^{}_{}[R_{i\alpha}, \dot{R}_{i\alpha}, \dot{B}_{\alpha}, \dot{C}] = 
2\int\d\lambda\sqrt{\fT^{}\{\fU + \fE\}} 
\eeq
with $\fT^{}(R_{i\alpha}, \dot{R}_{i\alpha}, \dot{B}_{\alpha}, \dot{C}) = M^{ij\alpha\beta}
\circ R_{i\alpha} \circ R_{j\beta}/2\mI$ for $\mI$ the total barycentric moment of inertia of the system, 
whereupon $\fV$ is a function only of manifestly scale-invariant {\sl ratios} of 
$\sqrt{\underline{R}_i\cdot\underline{R}_j}$ (i.e. homogeneous of degree zero) and $\fE$ comes unweighted.
[Barbour's original presentation \cite{B03}, in which the potential is homogeneous of degree $-2$ and 
an energy E cannot be added on to this (but $\fE/\mI$ {\sl can}), is related to this one by use of 
$\Omega^2 = \mI$, i.e. by not dividing through by the moment of inertia in the first place.]  
Then (in the conformally natural presentation I use above) the conjugate momenta are given by 
(\ref{Rmom}) but divided by $\mI$ and containing the $*$ corresponding to the new $\circ$.
There are again primary and secondary constraints respectively of form (\ref{En}) with the first term 
multiplied by $\mI$ and (\ref{ZAM}), as well as a secondary constraint from variation with respect to $C$, 
\beq
\ttD \equiv R_{i\alpha}P^{i\alpha} = 0 \mbox{ } , 
\label{ZDM}
\eeq
which is also linear in the momenta.  
This is the dilational (or Euler) constraint -- it says that the dilational momentum of the whole system 
is zero so that the physical content of the theory is not in relative separations but in ratios 
of relative separations (relative angles already being functions of ratios, they are unaffected).

\subsection{Relational formulation of geometrodynamics}

Important further motivation for the study of relationalism and relational particle mechanics is that 
General Relativity (GR) can also be formulated as a relational theory.
Begin by recollecting that as well as being a spacetime theory, GR can be studied as a dynamics obtained 
by splitting spacetime with respect to a family of spatial hypersurfaces \cite{ADM}.  
However, answering a question of Wheeler \cite{Battelle}, this dynamics can be taken to follow from 
first principles of its own \cite{HKT, RWR, Phan}.   
And one of the two known such sets of first principles are relational first principles, with $\fQ$ = 
Riem($\Sigma$, 3) -- the space of positive-definite 3-metrics on some fixed topology $\Sigma$ which I 
take to be a compact without boundary one for simplicity -- and $G$ = Diff($\Sigma$, 3) the diffeomorphisms on $\Sigma$ \cite{RWR, Lan, Phan}.   
The arbitrary Diff($\Sigma$, 3) frame expressions are 

\noindent
$\circ h_{\mu\nu} \equiv \dot{h}_{\mu\nu} - {\pounds}_{\dot{F}}h_{\mu\nu}$ 
rather than `bare' $\dot{h}_{\mu\nu}$.\footnote{The
dot now denotes ${\pa}/{\pa\lambda}$.  
$F_{\mu}$ generates Diff($\Sigma$, 3).
$\pounds_{\dot{F}}$ is the Lie derivative with respect to the vector field $\dot{F}_{\mu}$.
$h$, $D_{\mu}$ and $\mbox{Ric}(h)$ are the determinant, covariant derivative and Ricci scalar associated 
with $h_{\mu\nu}$.  
$\Lambda$ is the cosmological constant.  
${\cal M}^{\mu\nu\rho\sigma} = h^{\mu\rho}h^{\nu\sigma} - h^{\mu\nu}h^{\rho\sigma}$ is the kinetic 
supermetric of GR, with determinant ${\cal M}$ and inverse ${\cal N}_{\mu\nu\sigma\rho} = 
h_{\mu\rho}h_{\nu\sigma} - h_{\mu\nu}h_{\rho\sigma}/2$ (which is the undensitized version of the 
DeWitt supermetric \cite{DeWitt}).}

The action that one builds so as to explicitly implement temporal relationalism is \cite{RWR, Lan, ABFO, Phan} 
\beq
\fS^{\sG\sR}_{}[h_{\alpha\beta}, \dot{h}_{\alpha\beta}, \dot{F}_{\beta}] = 
2\int\d\lambda\int\d^{3}x\sqrt{h}\sqrt{\fT^{\sG\sR}_{}
\{\mbox{Ric}(h) - 2\Lambda\}} \mbox{ } \mbox{  for } \mbox{ }   
\fT^{\sG\sR}_{}(x^{\omega}, h_{\mu\nu}, \dot{h}_{\mu\nu}; \dot{F}_{\mu}] = 
\frac{1}{4}{\cal M}^{\mu\nu\rho\sigma}\circ h_{\mu\nu}\circ h_{\rho\sigma}
\label{GRaction} \mbox{ } , 
\eeq
which bears many similarities to the better-known Baierlein--Sharp--Wheeler \cite{BSW} 
action.\footnote{The
difference is that the Baierlein--Sharp--Wheeler action contains the shift, $\beta_{\mu}$, while the 
action (\ref{GRaction}) contains the velocity associated with the frame variable $F_{\mu}$, which is 
such that $\dot{F}_{\mu} = \beta_{\mu}$.
Both of these actions are free of extraneous time-related variables (unlike the Arnowitt--Deser--Misner 
action which contains a such -- the lapse -- multiplier elimination of which produces the 
Baierlein--Sharp--Wheeler action).
The above-mentioned difference does not affect the outcome of the variational procedure \cite{FEPI}; 
however, it does make (\ref{GRaction}) homogeneous quadratic in $\pa/\pa\lambda$ so that the $\lambda$'s 
cancel out of $\d\lambda\sqrt{\fT}$ so that it and not the Baierlein--Sharp--Wheeler action is 
manifestly reparametrization invariant as well as free of extraneous time-related variables, 
i.e. temporally relational as defined on page 1.}
%
Also note that one does not need to assume the GR form of the kinetic metric or potential; relational 
postulates plus a few simplicities give this since the Dirac procedure \cite{Dirac} prevents most other 
likewise simple choices of kinetic term $\fT$ from working \cite{RWR, SanOM, Than, Lan, Phan}.     
Yet further motivation is that 1) configurational relationalism is closely related \cite{Lan, FEPI} to 
certain formulations of gauge theory.  
2) The above relational formulation of GR is furthermore robust to the inclusion of a sufficiently full 
set of fundamental matter sources so as to describe nature \cite{AB, Van, Lan, Phan, Lan2}.

Then the conjugate momenta are 
\beq
\pi^{\mu\nu} = \sqrt{h}{\cal M}^{\mu\nu\rho\sigma}*h_{\rho\sigma} \mbox{ } \mbox{ for } \mbox{ }
*h_{\rho\sigma} = {h_{\rho\sigma}}^* - \pounds_{F^*}h_{\rho\sigma} 
\label{tumvel}
\eeq
for $\mbox{}^* \equiv \sqrt{\{\mbox{Ric}(h) - 2\Lambda\}/{\fT^{\sG\sR}_{}}}\mbox{ } \dot{\mbox{}} \mbox{ }$.
(\ref{GRaction}) being reparametrization invariant, there must likewise be at least one primary 
constraint, which is in this case the GR Hamiltonian constraint
\beq
{\cal H} \equiv \frac{1}{\sqrt{h}}{\cal N}_{\mu\nu\rho\sigma}\pi^{\mu\nu}\pi^{\rho\sigma} - 
\sqrt{h}\{ \mbox{Ric}(h) - 2\Lambda\} = 0  \mbox{ }   
\label{Ham}
\eeq
to which the momenta contribute quadratically but not linearly. 
Variation with respect to $F_{\mu}$ yields as a secondary constraint the GR momentum constraint
\beq
{\cal H}_{\mu} \equiv - 2D_{\nu}{\pi^{\nu}}_{\mu} = 0 \mbox{ } ,   
\label{Mom}
\eeq
which is linear in the gravitational momenta and interpretable as GR being more than just a theory of 
dynamical 3-metrics: the physical information is in the underlying geometry and not in the allocation 
of points to that geometry. 
While, the purely quadratic nature of ${\cal H}$ leads to $\widehat{\cal H}\Psi = \fE\Psi$ and not 
${i\hbar}{\pa\Psi}/{\pa T}$ for some notion of time $T$.  
The $\ttH$ of relational particle mechanics also has this feature and so itself manifests the Problem 
of Time (Sec II.1); it is a reasonable model for this in that a number of the conceptual strategies 
subsequently suggested for GR have nontrivial counterparts for relational particle mechanics.

The presence of square roots in the above actions so as to implement manifest temporal relationalism 
has caused some concern with Referees due to square roots causing substantial difficulties at the 
quantum level.  
In this respect, I note that the above square roots occur {\it at the classical level in the Lagrangian 
formulation}.  
However, {\sl there are no square roots in the Hamiltonians resulting from such actions}, 
and it is these that I promote to quantum equations in \cite{08II}, so that there are 
{\sl no square roots in the quantum equations}, and so this work encounters none of the problems 
associated with handling square roots at the quantum level.

\subsection{Relational formulation of conformogeometrodynamics}

This has a number of ties with the work of Lichnerowicz \cite{Lich44} and York \cite{York72} on maximal 
and constant mean curvature spatial slices respectively, which is important in numerical 
relativity \cite{CMCApps}.    
A preliminary action is 
\beq
\fS^{\sG\sR}_{} = 
\int\d\lambda\int\d^3x\sqrt{h}\phi^6\sqrt{\fT^{\sG\sR}_{}\{\phi^{-4}\{\mbox{Ric}(h) -8D^2\phi/\phi\}\}}
\label{GRconfaction} \mbox{ } , 
\eeq
where\foo{I, like \cite{BO}, 
present this for simplicity with no $\Lambda$ term; see \cite{ABFO} for inclusion of this.  
Also, \cite{BO} first considered this with two separate multipliers instead of a single more general 
auxiliary whose velocity also features in the action and has to be free end hypersurface varied; the way 
this is presented here is that of \cite{ABFKO} and \cite{FEPII}.   
See also \cite{FEPI, FEPII} for justification of the type of variation in use.}
$$
\fT^{\sG\sR}_{}(x^{\omega}, h_{\mu\nu}, \phi, \dot{h}_{\mu\nu}, \dot{\phi}_{\mu}; \dot{F}_{\mu}] = 
\{\phi^{-4}h^{\mu\rho}\phi^{-4}h^{\nu\sigma} - \phi^{-4}h^{\mu\nu}\phi^{-4}h^{\rho\sigma}\}
\circ\{\phi^4h_{\mu\nu}\}\circ\{\phi^4h_{\rho\sigma}\} = 
$$
\beq
\{h^{\mu\rho}h^{\nu\sigma} - h^{\mu\nu}h^{\rho\sigma}\}
\{\circ h_{\mu\nu} +     4 h_{\mu\nu}    \circ\phi/\phi\}
\{\circ h_{\rho\sigma} + 4 h_{\rho\sigma}\circ\phi/\phi\} \mbox{ } ;
\eeq
$\phi$ is a conformal factor.
This action then gives as a primary constraint 
\beq
{\cal H}_{\phi} \equiv \phi^{-8}\frac{1}{\sqrt{h}}\pi_{\mu\nu}\pi^{\mu\nu} - \sqrt{h}  
\left\{
\mbox{Ric}(h) - \frac{8D^2\phi}{\phi}
\right\} = 0 
\eeq
(Lichnerowicz equation), ${\cal H}_{\mu}$ as a secondary constraint from $F_{\mu}$ variation, and 
\beq
\pi = 0
\eeq 
(condition for a maximal slice) as one part of the free end hypersurface \cite{FEPI, FEPII} variation of 
$\phi$, but the other part of this variation entails frozenness (the well-known non-propagability of 
maximal slicing for spatially compact without boundary GR.

\cite{BO} and \cite{ABFO} got round this frozenness was circumvented by considering a new action with 
division by $\mbox{Vol}^{2/3}$ which amounts to fully using $G$ = Diff($\Sigma$, 3) $\times$ 
Conf($\Sigma$, 3) where Conf($\Sigma$, 3) is the group of conformal transformations on $\Sigma$.  
But the subsequent variational principle no longer gives GR and is furthermore questionable as an 
alternative theory \cite{ABFO, Lan, Than}).

In \cite{ABFKO} it was circumvented rather by using $G$ = Diff($\Sigma$, 3) $\times$ VPConf($\Sigma$, 3), 
where VPConf($\Sigma$, 3) are the (global) volume-preserving conformal transformations on $\Sigma$ as 
implemented by using 
\beq
\hat{\phi} = \phi/\left\{\int\d^3x\sqrt{h}\phi^6\right\}^{1/6} \mbox{ } .  
\eeq
Subsequently, the primary constraint (now denoted by ${\cal H}_{\hat{\phi}}$) picks up more terms (making it a Lichnerowicz-York equation rather than a 
Lichnerowicz equation).  
While, $\phi$ variation now gives 
\beq
\pi/\sqrt{h} = C \mbox{ } ,
\eeq
(condition for a constant mean curvature slice) alongside an equation that successfully maintains this.  
The addition of matter to the present paragraph's scheme has not to date been extensively 
studied, but no significant hindrances are known to date \cite{MacSweeney}, and the addition of matter 
to the preceding paragraph's scheme has been extensively studied \cite{ABFO, CMCAnderson}.   


\section{Geometry of the configuration spaces}

The following configuration space structure issues play an important underlying role in papers I and II.

\subsection{Euclidean relational particle mechanics configuration spaces and reduction} 

For Euclidean relational particle mechanics, such a study was carried out in \cite{GGM} in a `rigged' fashion (auxiliary variables 
included), and in \cite{LB, GGM, 06I} in a reduced fashion (auxiliary variables eliminated).   
\cite{RelatedWork} exemplifies related work.  
In \cite{06II} I noted that this elimination has a simpler nature in 2-d than in 3-d, and used this to 
pass to completely relational variables in \cite{TriCl, FORD} for both Euclidean and similarity relational 
particle mechanics.
{\it Relative space} $\fR(\mN, \d)$ is the quotient space $\fQ(\mN,\d)/\mbox{Tr}(\d)$.   
This is flat space, so the kinetic metric is just $\fT^{}(\underline{R_i}) = 
\sum_i\mu_iR_i^2/2$.  
At this level have Jacobi coordinates and constraints as in Sec 2.1.  
I use mass-weighted Jacobi variables as the most succinct `original variables of the problem'.  
{\it Relational space} is ${\cal R}(\mN, \d) = \fQ(\mN,\d)/\mbox{Eucl}(\d)$, which is what one has 
on eliminating the angular momentum constraint.  
These coincide in 1-d.  
In 2-d one can do elimination explicitly \cite{TriCl, FORD}, obtaining e.g. in the 3-particle case the 
$\fT^{\st\sr\si\sa\sn\sg\sll\se}(\iota_i, \dot{\iota}_i, \dot{\Phi})$ corresponding to the line element 
\beq
\d s^2_{\st\sr\si\sa\sn\sg\sll\se} =  \d{\iota}_1^2 + \d{\iota}_2^2  + 
\frac{\iota_1^2\iota_2^2\d {\Phi}^2}{\iota_1^2 + \iota_2^2}   
\mbox{ } .   
\eeq
[To obtain a configuration space kinetic term $\fT$ from the line element of a metric 
$\d s^2 = M_{\sfA\sfB}\d Q^{\sfA}\d Q^{\sfB}$, use (\ref{Te}).]  
This elimination is done by: Step 1) passing to mass-weighted Jacobi coordinates.  
Step 2: eliminating the rotational auxiliary from the Lagrangian form of the constraint (\ref{ZAM}), 
casting the subsequent expression in mass-weighted Jacobi bipolar coordinates, which I denote by 
($\iota_1$, $\theta_1$, $\iota_2$, $\theta_2$) [fig 1iv); these are still with respect to fixed axes.] 
Step 3: one can then pass to fully Euclideanly-relational coordinates [fig 1v)] ($\{\iota_1, \iota_2, 
\Phi\}$) (by an additional `purely absolute' angle dropping out of the working \cite{06I, TriCl}). 
In 3-d, the situation is considerably harder.

\subsection{Similarity relational particle mechanics configuration spaces and reduction } 

For similarity relational particle mechanics, such a study was carried out in \cite{FORD} in a reduced fashion (both by reduction and 
in an already-reduced form based on \cite{Kendall}).  
{\it Preshape space} is $\fP(\mN, \d) = \fQ(\mN,\d)/\mbox{Tr}(\d) \times \mbox{Dil}$ and {\it Shape 
space} is $\fS(\mN, \d) = \fQ(\mN,\d)/\mbox{Sim}(\d)$.  
Furthermore, $\fP(\mN, \d) {=} \mathbb{S}^{\sn\sd - 1}$ and $\fS(\mN, 1) 
{=} \mathbb{S}^{\sn - 1}$ where in both cases `=' here means equal both 
topologically and metrically, with the standard (hyper)spherical metric. 
Thus one has $\fT^{\sN\s-\sss\st\so\sp}({\cal R}_{\barp}, \dot{\cal R}_{\barp})$ or 
$\fT^{\sN\s-\sss\st\so\sp}(\Theta_{\barp}, \dot{\Theta}_{\barp})$ corresponding to the line element 
\beq
\d s^2_{\sN\s-\sss\st\so\sp} = \frac{         
\left\{     1 + \sum_{\barp = 1}^{\sn - 1}{\cal R}_{\barp}^2     \right\}
                     \sum_{\barq = 1}^{\sn - 1}\d{{\cal R}}_{\barq}^2 - 
                     \left\{     \sum_{\barp = 1}^{\sn - 1}{\cal R}_{\barp}\d{{\cal R}}_{\barp}    
\right\}^2       }   
           {                     \{1 + \sum_{\barp = 1}^{\sn - 1}{\cal R}_{\barp}^2\}^2                      } 
 = \sum_{\barr = 1}^{\sn - 1}
\prod_{\hat{p} = 1}^{\barr - 1}\mbox{sin}^2\Theta_{\hat{p} }\d{\Theta}_{\barr}^2 \mbox{ } 
\eeq
in terms of simple ratio coordinates (\ref{SRC}) and (ultra)spherical coordinates (\ref{Ultra}), 
respectively, where $\prod_{i = 1}^{0}$ terms are defined to be 1.  
It is obtained from the similarity relational particle mechanics action in Jacobi coordinates by: Step 1) passing to mass-weighted 
coordinates.
Step 2) eliminate dilational auxiliaries from the Lagrangian form of the constraint (\ref{ZDM}) and 
re-express in mass-weighted Jacobi bipolar coordinates. 
Step 3) pass to fully relational simple ratio variables 
\beq
{\cal R}_{\barp} \equiv \iota_{\barp}/\iota_{\sn - 1} 
\label{SRC}
\eeq
(in the present context these are, geometrically, Beltrami coordinates); these are related to 
ultraspherical coordinates by 
\beq
\Theta_{\barp} = \mbox{arctan}
\left(
{    \sqrt{\sum_{j = 1}^{\barp}{\cal R}_{j}^2}    }/{    {\cal R}_{\barp + 1}    }
\right)
\label{Ultra}
\eeq
for ${\cal R}_{\sn - 1} \equiv \iota_{\sn - 1}/\iota{\sn - 1} = 1$.
The coordinate ranges for these are $\Theta_{\barp} \in (0, \pi)$, $\Theta_{\sn\d - 1} \in [0, 2\pi)$.
E.g. for 4-stop metroland $\Theta_1$ is the azimuthal angle $\Theta$ and $\Theta_2$ is the polar angle $\Phi$.  
Then one has $\fT^{4\s-\sss\st\so\sp}({\cal R}_i, \dot{\cal R}_i)$ or 
$\fT^{\sN\s-\sss\st\so\sp}(\Theta, \dot{\Theta}, \dot{\Phi})$ corresponding to the line element 
\beq
\d s^2_{4\s-\sss\st\so\sp} = 
\frac{\{1 + {\cal R}_2^2\}\d{{\cal R}}_1^2 + \{1 + {\cal R}_1^2\}\d{{\cal R}}_2^2 - 
2{\cal R}_1{\cal R}_2\d{{\cal R}}_1\d{{\cal R}}_2}{\{1 + {\cal R}_1^2 + {\cal R}_2^2\}^2} =
\d{\Theta}^2 + \mbox{sin}^2\Theta\d{\Phi}^2 
\eeq
for `simple ratio' coordinates (${\cal R}_1$, ${\cal R}_2$) [4-stop metroland subcase of (\ref{SRC})] 
and spherical coordinates ($\Theta, \Phi$) [ = ($\Theta_1, \Theta_2$), given by the 4-stop metroland  
subcase of (\ref{Ultra})].

While, for N particles in 2-d $\fS(\mN, 2) {=} \mathbb{CP}^{\sn - 1}$ both topologically and metrically, 
with the standard Fubini--Study metric corresponding to the line element 
\beq
\d s^2_{\sN\s-\sa\s-\sg\so\sn} =  
\frac{    \{1 + \sum_{\barp}|{\cal Z}_{\barp}|^2\} 
                \sum_{\barq}|\d{\cal Z}_{\barq}|^2 - 
                 |\sum_{\barp}\overline{{\cal Z}}_{\barp} \d{\cal Z}_{\barp}|^2    }
            {    \{1 + \sum_{\barr}|{\cal Z}_{\barr}|^2\}^2    } \mbox{ } .
\label{10}
\eeq
for ${\cal Z}_{\barr}$ inhomogeneous coordinates on $\mathbb{CP}^{\sn - 1}$, from which the kinetic term 
$\fT^{\sN\s-\sa\s-\sg\so\sn} ({\cal Z}_{\barp}, \dot{\cal Z}_{\barp})$ is constructed.
This is obtained from the similarity relational particle mechanics action in Jacobi coordinates by Step 1) passing to mass-weighted 
coordinates.
Step 2) eliminate both a rotational auxiliary and a dilational auxiliary from the 
Lagrangian forms of (\ref{ZAM}, \ref{ZDM}). 
Step 3) write this action in terms of ratio variables.  
Furthermore, one can use the polar form ${\cal Z} = {\cal R}_{\barp}\mbox{exp}(i\Theta_{\barp})$;      
indexing moduli (real ratio coordinates)     as ${\cal R}_{\barp}$ 
and arguments   (relative angle coordinates) as ${\Theta}_{\tilde{p}}$ 
(for both $\barp$ and $\tilde{p}$ taking 1 to N -- 2).
Then the configuration space metric can be written in two blocks (${\cal M}_{\barp\tiq} = 0$):
\beq
{\cal M}_{\barp\barq} = \{1 + ||{\cal R}||^2\}^{-1}\delta_{\barp\barq} - 
\{1 + ||{\cal R}||^2\}^{-2}{\cal R}_{\barp}{\cal R}_{\barq} 
\mbox{ } , \mbox{ }
{\cal M}_{\tip\tiq} = \{\{1 + ||{\cal R}||^2\}^{-1}\delta_{\tip\tiq} - 
\{1 + ||{\cal R}||^2\}^{-2}{\cal R}_{\tip}{\cal R}_{\tiq}\} {\cal R}_{\tip}{\cal R}_{\tiq}
\mbox{ } \mbox{   (no sum) ,}
\eeq 
where for a given p ${\cal R}_{\tilde{p}}$ is the ${\cal R}_{\bar{p}}$ that forms a complex coordinate 
pair with ${\Theta}_{\tilde{p}}$.

As a particular example, $\fS$(3, 2) = $\mathbb{CP}^1 = \mathbb{S}^2$, making this triangleland case 
particularly amenable to study due to the availability of both projective and spherical techniques 
(while, as we shall see, this example meets many of the nontrivialities required by Problem of Time 
strategies); this paper principally considers this case.  
In this case the kinetic term collapses to $\fT^{\st\sr\si\sa\sn\sg\sll\se}({\cal Z}, \dot{\cal Z})$, 
$\fT^{\st\sr\si\sa\sn\sg\sll\se}_{\mathbb{CP}^1}({\cal R}, \dot{\cal R}, \dot{\Phi})$ or
$\fT^{\st\sr\si\sa\sn\sg\sll\se}_{(\mathbb{S}^2, 1/2)}(\Theta, \dot{\Theta}, \dot{\Phi})$ as constructed from 
the line element 
\beq
\d s_{\st\sr\si\sa\sn\sg\sll\se}^2 = {|\d{\cal Z}|^2}/{\{1 + |{\cal Z}|^2\}^2} = 
\{\d{\cal R}^2 + {\cal R}^2\d{\Phi^2}\}/{\{1 + {\cal R}^2\}^2} = 
\{\d{\Theta}^2 + \mbox{sin}^2\Theta \d{\Phi}^2\}/4 \mbox{ } ,
\label{T32}
\eeq
where ${\cal R}$ is the simple ratio variable ${\cal R} \equiv {\iota_1}/{\iota_2}$ 
This is physically the square root of the ratio of partial moments of inertia, $\sqrt{\mI_1/\mI_2}$, 
and mathematically a choice of inhomogeneous coordinate's modulus on $\mathbb{CP}^1$.
This is related to the azimuthal angle $\Theta$ on $\mathbb{S}^2 = \mathbb{CP}^1$ by 
\beq
\Theta = 2\mbox{arctan}{\cal R} \mbox{ } \Leftrightarrow \mbox{ } {\cal R} = \mbox{tan$\frac{\Theta}{2}$}
\eeq 
for $\Theta$ the spherical azimuthal angle (see also Fig 2).  
[Thus ${\cal R}$ is also geometrically interpretable as a radial stereographic coordinate on 
$\mathbb{S}^2$].    
$\Phi$ is the relative angle between the two Jacobi coordinate vectors.
Useful relations between the two representations include that ${\cal R} = 1$ is the equator,   
${\cal R}$ small (compared to 1) corresponds to near the North Pole ($\Theta$ a small angle)
and ${\cal R}$ large corresponds to near the South Pole (the supplement angle 
$\Xi = \pi - \Theta$ is small).
Also, one can use the barred banal conformal representation $\fT \longrightarrow 4\fT$, 
$\fU + \fE \longrightarrow 4\{\fU + \fE\}$, whereupon this sphere of radius 1/2 becomes the unit sphere, 
and I write $\overline{\fT}_{\mathbb{S}^2}^{\st\sr\si\sa\sn\sg\sll\se}$.    

\mbox{ }

\noindent {\footnotesize[{\bf Figure 2}: Interrelation between ${\cal R}$, ${\cal U}$, $\Theta$ and 
$\Xi$ coordinates.
N is the North Pole, S is the South Pole, O is the centre.  
Then point P in the spherical representation corresponds to point $P^{\prime}$ in the stereographic 
tangent plane at N with radial coordinate ${\cal R}$, and to point $P^{\prime\prime}$ in the 
stereographic tangent plane at S with radial coordinate ${\cal U}$.]}
  
\mbox{ } 

For use in paper II, this line element in terms of $\Phi$ and $\mI_i$, $i = 1$ or $2$ is (no sum) 
\beq
\d s^{2} = {\d\mI_i^2}/{4\mI_i\{\mI - \mI_i\}} + {\{\mI - \mI_i\}\mI_i}\d\Phi^2/\mI^2 \mbox{ } .
\eeq

One can also use ${\fT}_{\sf\sll\sa\st} \equiv 
\widetilde{\fT}^{\st\sr\si\sa\sn\sg\sll\se}({\cal R}, \dot{\cal R}, \dot{\Phi})$ constructed from the line 
element 
\beq
\d s^2_{\sf\sll\sa\st} = \dot{\cal R}^2 + {\cal R}^2\dot{\Phi}^2  
\label{flatbrod}
\eeq
by performing a banal conformal transformation with conformal factor $\Omega^2 = \{1 + {\cal R}^2\}^2$.  
This is geometrically trivial, while the other above forms are both geometrically natural and 
mechanically natural (equivalent to $\fE$ appearing as an eigenvalue free of weight function); 
using $\{1 + {\cal R}^2\}$ alone would be the conformally-natural choice.

In 3-d, the situation is, again, harder \cite{Kendall} though at least similarity relational particle mechanics is free of 
collisions that are {\sl maximal} (i.e. between all the particles at once).

\subsection{Euclidean relational particle mechanics in scale--shape variables} 

Finally, these `shapes' are also relevant within the corresponding Euclidean relational particle 
mechanics, as these admit conceptually interesting formulations in terms of scale--shape split variables.  
For scaled N-stop metroland, the configuration space is the generalized cone $C(\mathbb{S}^{\sn - 1})$ 
while for N-a-gonland  it is $C(\mathbb{CP}^{\sn - 1})$, scaled triangleland also being 
$C(\mathbb{S}^2)$.  
The special cone-point or `apex' 0 here corresponds physically to the maximal collision.  
Now use the new coordinate 
\beq
\mI = \iota_1^2 + \iota_2^2  \mbox{ } , \mbox{ }
\eeq
which is physically the moment of inertia and mathematically a radius, 
and the same ${\cal R}$ coordinate as in Sec 3.2.   
This coordinate transformation then inverts to 
\beq
\iota_1 = \sqrt{{\mI}/\{1 + {\cal R}^2\}}{\cal R} \mbox{ } , \mbox{ } 
\iota_2 = \sqrt{{\mI}/\{1 + {\cal R}^2\}} \mbox{ } .  
\eeq
Also, ${\cal R}$ can be supplanted by $\Theta = 2\mbox{arctan}{\cal R}$.

The kinetic term is $\fT^{\st\sr\si\sa\sn\sg\sll\se}(\mI, {\cal Z}, \dot{\mI}, \dot{\cal Z})$, 
$\fT^{\st\sr\si\sa\sn\sg\sll\se}(\mI, {\cal R}, \dot{\mI}, \dot{\cal R}, \dot{\Phi})$ or
$\fT^{\st\sr\si\sa\sn\sg\sll\se}(\mI, \Theta, \dot{\mI}, \dot{\Theta}, \dot{\Phi})$ as constructed from the line 
element
\beq
\d s^2_{\st\sr\si\sa\sn\sg\sll\se}
 = \frac{1}{4\mI}
\left\{ 
\d{\mI}^2 + 4\mI^2\frac{|\d{\cal Z}^2|}{\{1 + |{\cal Z}|^2\}^2}
\right\}
 = \frac{1}{4\mI}
\left\{ 
\d{\mI}^2 + 4\mI^2\frac{\d{\cal R}^2 + {\cal R}^2\d{\Phi}^2}{\{1 + {\cal R}^2\}^2}
\right\}
 = \frac{1}{4\mI}
\left\{ 
\d{\mI}^2 + \mI^2\{\d{\Theta^2} + \mbox{sin}^2\Theta\d{\Phi^2}\}
\right\} \mbox{ } .  
\eeq
Moreover, one can use instead the banal-conformally related kinetic term 
$\check{\fT}_{\sF\sll\sa\st}(\mI, \Theta, \dot{\mI}, \dot{\Theta}, \dot{\Phi})$ constructed from 
\beq
\d s^2_{\sF\sll\sa\st} = 
\d{\mI}^2 + \mI^2\{ \d{\Theta}^2 + \mbox{sin}^2\Theta\d{\Phi}^2 \}
\label{fltrp} \mbox{ } 
\eeq
(the corresponding conformal factor being $4\mI$, so that also 
\beq
\left.
\check{\fU} + \check{\fE} = \{\fU + \fE\}/4\mI \mbox{ } 
\right) \mbox{ } ,   
\eeq
away from $\mI = 0$ in which place this conformal transformation is invalid).  
The form of (\ref{fltrp}) clearly makes this a flat (i.e. geometrically trivial) representation in 
spherical polar coordinates, with $\mI$ as the radius.  
While, the other forms above are mechanically natural.\footnote{\cite{Cones} 
connects this section's workings with results from the celestial mechanics and molecular physics 
literatures.}

\subsection{Geometrodynamical configuration spaces} 

While, GR is a {\it geometrodynamics} \cite{Battelle, DeWitt} on the quotient configuration space 
Superspace($\Sigma$, 3) $\equiv$ 

\noindent Riem($\Sigma$, 3)/Diff($\Sigma$, 3) \cite{Battelle, DeWitt}, 
which is studied topologically and geometrically in e.g. \cite{Battelle, DeWitt, Superspace}.  
Superspace is an infinite-dimensional complicatedly stratified manifold; explicit reduction is not in 
general possible here.

\subsection{Conformogeometrodynamical configuration spaces} 

One can obtain relational theories or formulations using G = Diff($\Sigma$, 3) $\times$ Conf($\Sigma$, 3) 
that reproduces the maximal condition (however this formulation freezes unless one alters one's theory 
from GR \cite{ABFO}) and using G = Diff($\Sigma$, 3) $\times$ VPConf($\Sigma$, 3) (the volume-preserving 
conformal transformations) that does reproduce GR in the York formulation from an action principle 
\cite{ABFKO}.

Here the associated configuration spaces are conformal superspace CS($\Sigma$, 3) $\equiv$ Riem($\Sigma$, 3)
/Diff($\Sigma$, 3) $\times$ Conf($\Sigma$, 3), which is studied geometrically in \cite{CS, FM96}, and 
\{CS + V\}($\Sigma$, 3) = Riem($\Sigma$, 3)/Diff($\Sigma$, 3) $\times$ VPConf($\Sigma$, 3), which 
previously featured in e.g. \cite{York72}, but has not to my knowledge been studied from a geometrical 
perspective.
[The CRiem($\Sigma$, 3) $\equiv$ Riem($\Sigma$, 3)/Conf($\Sigma$, 3) analogue of 
preshape space has been studied geometrically in e.g. \cite{DeWitt, FM96} ].

Parallels between this and Euclidean relational particle mechanics in shape-scale split variables 
include homogeneous degrees of freedom playing a similar role to scale, York time \cite{YorkTime} 
having an `Euler time' analogue \cite{06II}, similarity relational particle mechanics  
corresponding to maximal slicing in having this time frozen, and in the analogy between 
\{CS + V\}($\Sigma$, 3) and the Euclidean relational particle mechanics configuration spaces' cone 
structure.    
In the configuration space of GR, the analogous `special point' at zero scale is the Big Bang.

\section{Similarity relational particle mechanics at the classical level}

\subsection{1 and 2-d cases} 

In App A, I provide how the general curved space mechanics unfolds.  
In the case of scalefree N-stop metroland, the Jacobi action is
\beq
\fS_{}^{\sN\s-\sss\st\so\sp}(\Theta_{\barp}, \dot{\Theta}_{\barp}) = 
2\int\d\lambda\sqrt{\fT_{\mathbb{S}^{\tn - 1}}\{\fE - \fV\}} \mbox{ } .   
\eeq
The Euler--Lagrange equations following from this are given in \cite{FORD} in Beltrami coordinates (the notation there uses 
$s_{\barp}$ in place of the present paper's ${\cal R}_{\barp}$), but in this paper I use, rather, 
(ultra)spherical coordinates in the present paper, so I provide the Euler--Lagrange equations afresh in these: 
\beq
\left\{\prod_{\barp}^{\barq - 1} \mbox{sin}^2\Theta_{\barp} \Theta_{\barq}^*\right\}^* - 
\left\{
\sum_{\barr = \barq + 1}^{\sn - 1} 
\prod_{\barp = 1, \barp \neq \barq}^{\barr - 1}\mbox{sin}^2\Theta_{\barp}
\right\} 
\mbox{sin}\Theta_{\barq}\mbox{cos}\Theta_{\barq}\Theta_{\barr}^{*2} = - \frac{\pa\fV}{\pa\Theta_{\barq}}
\mbox{ } .  
\eeq
There is also a first energy integral,  
\beq
{\sum_{\barr = 1}^{\sn - 1}
      \prod_{\barp = 1}^{\barr - 1}\mbox{sin}^2\Theta_{\barp}{\Theta}_{\barr}^{*2}}/{2} 
+ \fV(\Theta_{\barr}) = \fE \mbox{ } .  
\eeq
%

For scalefree N-a-gonland, the Jacobi action 
\beq
\fS = 2\int\d\lambda\sqrt{\fT_{\sF\sS}\{\fE - \fV\}} 
\eeq
gives the Euler--Lagrange equations as presented implicitly in \cite{FORD}.

In the exceptional triangleland case, there are spherical flat (conformal-to-stereographic) presentations,

\noindent
\beq
\fS_{}  = 2\int\d\lambda\sqrt{\overline{\fT}_{\mathbb{S}^2}\{\overline{\fE} - \overline{\fV}\}} 
           = 2\int\d\lambda\sqrt{\tilde{\fT}_{\sf\sll\sa\st}\{\widetilde{\fE} - \widetilde{\fV}\}} 
\mbox{ } .     
\eeq
Then, in the spherical presentation, the Euler--Lagrange equations simplify to

\beq
\Theta^{\overline{**}} - \mbox{sin}\Theta\mbox{cos}\Phi \Phi^{\overline{*}2} = 
- \frac{\pa\overline{\fV}}{\pa\Theta} \mbox{ } \mbox{ } , \mbox{ } \mbox{ } 
\{\mbox{sin}^2\Theta \Phi^{\overline{*}}\}^{\overline{*}} = - \frac{\pa\overline{\fV}}{\pa\Phi}
\label{Su}
\eeq
with an accompanying energy integral 
\beq
{\Theta^{\overline{*}2}}/{2} + {\mbox{sin}^2\Theta \Phi^{\overline{*}2}}/{2} + \overline{\fV} = 
\overline{\fE}  
\mbox{ } .  
\label{EN_FULL}
\eeq
[Above, 
\beq
\overline{*} \equiv \sqrt{\{\overline{\fE} + \overline{\fU}\}/{\overline{\fT}}}\mbox{ }\dot{} = 
\sqrt{\{\fE + \fU\}/{\fT}}/4\mbox{ }\dot{} = */4 \mbox{ } .] 
\label{overlinestardef} 
\eeq

If $\overline{\fV}$ is independent of $\Phi$, then (\ref{Su}ii) becomes another first integral:
\beq
\mbox{sin}^2\Theta \Phi^{\overline{*}} = {\cal J} \mbox{ } , \mbox{ constant }.  
\label{CF}
\eeq
This is a relative angular momentum quantity; see App C for various interpretations of it.  
Then (\ref{EN_FULL}) becomes 
\beq
{\Theta^{\overline{*}2}}/{2} + {{\cal J}^2}/{2\mbox{sin}^2\Theta} + 
\overline{\fV}(\Theta) = \overline{\fE}  
\mbox{ } . 
\label{MPW}
\eeq
Or, in terms of the ${\cal R}$ coordinate, the important results (\ref{CF}) and (\ref{MPW}) take the form 
\beq
{\cal R}^2\Phi^{\tilde{*}} = {\cal J} \mbox{ } , \mbox{ } 
\frac{1}{2}
\left\{
\left\{
\frac{\d {\cal R}}{\d \tilde{t}}
\right\}^2
+ \frac{{\cal J}^2}{{\cal R}^2} 
\right\} = \widetilde{\fE} + \widetilde{\fU}
\mbox{ } , 
\eeq
where $\tilde{*} \equiv \sqrt{\{\widetilde{\fE} + \widetilde{\fU}\}/\widetilde{\fT}}\dot{\mbox{ }}$.
This is illuminating though its direct parallel with the usual flat planar presentation 
of the ordinary mechanics of a test particle moving in a central potential: 
\beq
\mbox{\Huge(}
\stackrel{\mbox{ratio of square roots of the two subsystems'}}
         {\mbox{Jacobi partial moments of inertia }}  
\mbox{\Huge)} = {\cal R} \mbox{ } \longleftrightarrow \mbox{ } r = 
\mbox{ (radial coordinate of test particle) } , 
\eeq
\beq
\mbox{ (relative angle between Jacobi coordinates) } = \Phi \mbox{ } \longleftrightarrow \mbox{ } 
\theta = \mbox{ (polar coordinate of test particle) } , 
\eeq
\beq
\left(
\stackrel{\mbox{relative angular momentum}}{\mbox{of the two subsystems}}
\right) = {\cal J} \mbox{ } \longleftrightarrow \mbox{ } 
\mL \mbox{ } ( \mbox{ } = \mL_z \mbox{ } ) \mbox{ } =   
\left( 
\stackrel{\mbox{angular momentum component}}{\mbox{perpendicular to the plane}} 
\right) 
\mbox{ } ,
\eeq
\beq
1 \mbox{ } \longleftrightarrow \mbox{ } m = \mbox{ (test particle mass) }  
\mbox{ } .  
\eeq
This analogy is furthermore a pointer to parallel the well-known $u = 1/r$ substitution by the  
${\cal U} = 1/{\cal R}$ one, which turns out to be exceedingly useful in my study below.  
[In the spherical presentation the counterpart of this {\it inversion map}  
${\cal R} \longrightarrow {\cal U} = 1/{\cal R}$ is the {\it supplementary map} 
$\Theta \longrightarrow \pi - \Theta =$ the supplementary angle, $\Xi$.   
In $\iota_i$ or $I_i$ variables, it takes the form of interchanging the 1-indices and the 2-indices.
I term the underlying operation that takes these forms in these presentations as the {\it duality map}.]

Next, $\widetilde{\fV}_{\se\sf\sf} \equiv \widetilde{\fV} + {{\cal J}^2}/{{\cal R}^2} 
- \widetilde{\fE}$ is the potential quantity that is significant for motion in time, and combining 
(\ref{MPW}) and (\ref{CF}), $\widetilde{\fU}_{\so\sr\sb} \equiv - {\cal R}^4\widetilde{\fV}_{\se\sf\sf}$ 
is the potential quantity that is significant as regards the shapes of the classical orbits.  
%
%
%
Translating into the spherical language, the corresponding quantities are $\overline{\fV}_{\se\sf\sf} = 
\overline{\fV} + {J^2}/{\mbox{sin}^2\Theta} - \overline{\fE}$ and 
$\overline{\fU}_{\so\sr\sb} = - \mbox{sin}^4{\Theta}\overline{\fV}_{\se\sf\sf}$.

Finally, combining (\ref{MPW}) and (\ref{CF}) or their ${\cal R}$-analogues gives as quadratures for 
the shapes of the orbits 
\beq
\Phi - \Phi_0 = \int
{{\cal J}\d\Theta}/
     {\mbox{sin}\Theta\sqrt{2\{\overline{\fE} - \overline{\fV}(\Theta)\}\mbox{sin}^2\Theta - {\cal J}^2}}  
= \int {{\cal J}\d{\cal R}}/{{\cal R}\sqrt{2\{\widetilde{\fE} + 
\widetilde{\fU}\}{\cal R}^2 - {\cal J}^2}}
\mbox{ } .  
\label{quad}
\eeq
While, (\ref{MPW}) and its ${\cal R}$ analogue give as quadratures for the time-traversals
\beq
\overline{t} - \overline{t}_0 = \int {\mbox{sin}\Theta \d\Theta}/{\sqrt{2\{\overline{\fE} + 
\overline{\fU}\}\mbox{sin}^2\Theta - {\cal J}^2}} \mbox{ } , \mbox{ } \mbox{ or } \mbox{ }  
\widetilde{t} - \widetilde{t}_0 = \int {{\cal R}\d{\cal R}}/{\sqrt{2\{\widetilde{\fE} + 
\widetilde{\fU}\}{\cal R}^2 - {\cal J}^2}} \mbox{ } .  
\eeq

\subsection{A class of separable potentials for scalefree triangleland}

As $\theta$-independent (central) potentials considerably simplify classical and quantum mechanics, 
by the above analogy it is clear that $\Phi$-independent (relative angle independent) potentials  
will also considerably simplify classical and quantum scalefree triangleland.  
In particular, these simplifications include separability.  
Also, in each case, there is a {\sl conserved quantity}: angular momentum $\mL$ in the case of ordinary mechanics with 
central potentials, and {\sl relative} angular momentum ${\cal J}$ of the two constituent subsystems in the present 
context (see App C for details of the interpretation of ${\cal J}$).  
A general class of separable potentials consists of linear combinations of   
$$
\fV_{(\alpha, \beta)} 
\propto \left\{\frac{\iota_1}{\sqrt{\mI}}\right\}^{\alpha - \beta}
        \left\{\frac{\iota_2}{\sqrt{\mI}}\right\}^{\beta}
\propto \left\{\frac{\cal R}{\sqrt{ 1 + {\cal R}^2}}\right\}^{\alpha - \beta} 
        \left\{\frac{1}{\sqrt{ 1 + {\cal R}^2}}\right\}^{\beta} 
\propto \mbox{sin}^{\alpha - \beta}\mbox{$\frac{\Theta}{2}$}\mbox{cos}^{\beta}\mbox{$\frac{\Theta}{2}$}
$$
so that
\beq
\widetilde{\fV}_{(\alpha, \beta)} 
\propto {{\cal R}^{\alpha - \beta}}/{\{1 + {\cal R}^2\}^{\frac{\alpha}{2} + 2}}
\mbox{ } \mbox{ and } \mbox{ }
\overline{\fV}_{(\alpha, \beta)} 
\propto \mbox{sin}^{\alpha - \beta}\mbox{$\frac{\Theta}{2}$}\mbox{cos}^{\beta}\mbox{$\frac{\Theta}{2}$} 
\mbox{ } .  
\eeq
The most physically relevant subcase therein are the power-law-mimicking potentials $\beta = 0$  
(remember that $\mI$ turns out to be constant in similarity relational particle mechanics after 
variation is done).    
These correspond to potential contributions solely between particles 2, 3.  
The case $\beta = \alpha$ are potential contributions solely between particle 1 and the centre of mass 
of particles 2, 3 (which is less widely physically meaningful). 
It also turns out that the duality map sends $\fV_{(\alpha,\beta)}$ to $\fV_{(\alpha,\alpha - \beta)}$.

Simple examples therein of quantum-mechanical interest are the constant potential ($\alpha = 0$) and 
similarity relational particle mechanics' mimicker of harmonic oscillators ($\alpha = 2$).  
I choose these for explicit study due to the ubiquity of constant and harmonic oscillator potentials in theoretical 
physics; moreover harmonic oscillator potentials confer nice boundedness features at the quantum level.     
I use the following notation for the constants of proportionality for the single harmonic oscillator between particles 2, 
3:  
$2\fV_{(2, 0)}^{(1)} = h_{23}\{\underline{q}_2 - \underline{q}_3\}^2 \equiv H_1^{(1)}\rho_1^2$ for 
$h_{23}$ the ordinary position space Hooke's coefficient and $H_1$ the relative configuration space's 
`Jacobi--Hooke' coefficient.
I use the obvious cyclic permutations of this for harmonic oscillators between particles 3, 1, and particles 1, 2.  
Moreover, the linear combination of these last two such that $m_2h_{13} = m_3h_{12}$ (resultant force of 
the second and third `springs' pointing along the line joining the centre of mass of particles 2, 3 and 
the position of particle 1) is 
$2\{\fV_{(2, 0)}^{(2)} + \fV_{2, 2}^{(2)}\} = H_1^{(2)}\rho_1^2 + H_2^{(2)}\rho_2^2$, where 
\beq
H_1^{(2)} = \{h_{13}m_2^2 + h_{12}m_3^2\}/{\{m_{2} + m_{3}\}^2} \mbox{ } \mbox{ and } \mbox{ }   
H_2^{(2)} = h_{12} + h_{13} \mbox{ } .  
\label{H1and2}
\eeq
Additionally, one can consider $\fV = \fV^{(1)} + \fV^{(2)}$, whereupon one has 
$H_1 = H_1^{(1)} + H_1^{(2)}$ and $H_2 = H_2^{(2)}$; then define $K_i \equiv H_i/\mu_i$ so that 
\beq
2\fV = H_1\rho_1^2 + H_2\rho_2^2 = K_1\iota_1^2 + K_2\iota_2^2 \mbox{ } .  
\eeq
I refer to this case as the `special' case of the 3 harmonic oscillator problem, the general case having 
angle-dependent potentials (see the next SSec) due to {\sl not} having the resultant force 
of the second and third `springs' point along $\underline{\iota}_2$).  
Three notes of importance below and paper II are as follows.    

\noindent 1) In ($\cal R$, $\Phi$) coordinates this leads to the form 
\beq
2\widetilde{\fV} = \{K_1{\cal R}^2 + K_2\}/\{1 + {\cal R}^2\}^3 \mbox{ } ,
\eeq
or, in ($\Theta$, $\Phi$) coordinates, by the simplifications 
$\mbox{cos}^{2}\mbox{$\frac{\Theta}{2}$} = \{1 + \mbox{cos}\Theta\}/2$ and 
$\mbox{sin}^{2}\mbox{$\frac{\Theta}{2}$} = \{1 - \mbox{cos}\Theta\}/2$, to the form 
\beq
\overline{\fV} = A + B \mbox{cos}\Theta \mbox{ } \mbox{ for } \mbox{ }
A = 
\left\{
{K_1} + {K_2} 
\right\}/16
\mbox{ } , \mbox{ }
B = 
\left\{
{K_2} - {K_1} 
\right\}/16
\eeq
2) This situation is the linear combination of a $\fV_{(2, 0)}$ and a $\fV_{(2, 2)}$ and as such {\sl is 
self-dual provided that} the values of $K_1$ {\sl and} $K_2$ are also interchanged (or, equivalently, 
the sign of $B$ is reversed).    

\noindent 3) if $K_1 = K_2$ (corresponding to the highly symmetric balance 
$m_1h_{23} = m_2h_{31} = m_3h_{12}$ in position space), then 
\beq
\overline{\fV} = A  \mbox{ } , 
\eeq
to which I refer as the `very special case'; this is {\sl unconditionally self-dual}.

\subsection{Scalefree triangleland with harmonic oscillator type potential}

In all other multiple harmonic oscillator cases, there is a $\Phi$-dependent cross-term, rendering the potential 
nonseparable in these coordinates. 
The general triple harmonic oscillator like dipotential  $2\fV = h_{23}\{\q_2 - \q_3\}^2 +$ cycles maps to 
\beq
2\widetilde{\fV} = \{K_1{\cal R}^2 + L{\cal R}\mbox{cos}\Phi + K_2\}/\{1 + {\cal R}^2\}^3 
\mbox{ } \mbox{ for } \mbox{ } 
L \equiv \{\mu_1\mu_2\}^{-1/2}2\{h_{13}m_2 - h_{12}m_3\}/\{m_2 + m_3\} \mbox{ } .  
\eeq
Or, alternatively, in the spherical presentation, the potential is
\beq
\overline{\fV} = A + B\mbox{cos}\Theta + C\mbox{sin}\Theta\mbox{cos}\Phi
\mbox{ } \mbox{ for } \mbox{ } 
C = {L}/{16}
\mbox{ } . 
\eeq
This clearly includes the previous SSec's special case by setting $C = 0$ (and so $L = 0$, 
corresponding to $m_2h_{13} = m_3h_{12}$).

\subsection{Analogy with some linear rigid rotor set-ups}

Some useful mathematical analogies for scalefree triangleland with multiple harmonic oscillator like potentials are as 
follows.  
\beq
\mbox{ very special harmonic oscillator } \longleftrightarrow \mbox{ linear rigid rotor } ,
\eeq
\beq
\mbox{ special harmonic oscillator } \longleftrightarrow 
\mbox{linear rigid rotor in a background homogeneous electric field in the axial (`z')-direction } , 
\eeq
\beq
\mbox{ general harmonic oscillator } \longleftrightarrow 
\mbox{ linear rigid rotor in a background homogeneous electric field in an arbitrary direction }. 
\eeq
In particular, this classical problem has 
\beq
\fT_{\sr\so\st\so\sr} = \mI_{\sr\so\st\so\sr}\{\dot{\theta}^2 + \mbox{sin}^2\theta\dot{\phi}^2\}/{2} 
\mbox{ } , \mbox{ } \fV_{\sr\so\st\so\sr} = - {\cal M}{\cal E}\mbox{cos}\theta
\eeq
where $\mI_{\sr\so\st\so\sr}$ is the single nontrivial value of the moment of inertia of the linear 
rigid rotor, ${\cal E}$ is a constant external electric field in the axial `$z$' direction and 
${\cal M}$ is the dipole moment component in that direction.  
Thus the correspondence is $\Theta \longleftrightarrow \theta$, $\Phi \longleftrightarrow \phi$,  
\beq
1 \mbox{ } \longleftrightarrow \mbox{ } \mI_{\sr\so\st\so\sr} \mbox{ } , \mbox{ } 
\eeq
\beq
\mbox{(energy)/4 -- (sum of mass-weighted Jacobi--Hooke coefficients)/16 } 
\mbox{ } = \mbox{ }  
{\fE}/{4} - A 
\mbox{ } = \mbox{ } 
\overline{\fE} - A \mbox{ } \longleftrightarrow \mbox{ } E \mbox{ = (energy) } ,
\eeq
\beq
\mbox{ (difference of mass-weighted Jacobi--Hooke coefficients) = } \mbox{ } 
B \mbox{ } \longleftrightarrow 
\mbox{ } - {\cal ME} \mbox{ } , \mbox{ } 
\eeq
These all being well-studied at the quantum level \cite{TSMessiah, Hecht}, this identification is of 
considerable value in solving the relational problem in hand, by the string of techniques in Fig 3.  

\mbox{ }

\noindent{\footnotesize[{\bf Figure 3} In this paper I perform coordinate transformations to obtain  
standardly solvable problems and then back-track to see what happens in terms of the original 
mechanically-significant variables.]}

\subsection{A brief study of the potential}

Working in spherical coordinates, set $0 = {\pa\overline{\fV}}/{\pa\Theta} = - B\mbox{sin}\Theta + 
C\mbox{cos}\Theta\mbox{cos}\Phi$, $0 = {\pa\overline{\fV}}/{\pa\Phi} = -C\mbox{sin}\Theta\mbox{sin}\Phi$
to find the critical points.
These are at $(\Theta, \Phi) = (\mbox{arctan}(C/B), 0)$,  $(-\mbox{arctan}(C/B), \pi)$ which are 
antipodal (see Fig 4); in fact the potential is axisymmetric about the axis these lie on. 
The critical points are, respectively, a maximum and a minimum.  
[The very special case $B = C = 0$ is also critical, for all angles -- this case ceases to have a 
preferred axis.]


\mbox{ } 

\noindent{\footnotesize[{\bf Figure 4}]}

\mbox{ }

\noindent ${\cal R}$ small corresponds to $\Theta$ small, for which 
\beq
\overline{\fU} + \overline{\fE} = \{\overline{\fE} - A - B\} - C\Theta\mbox{cos}\Phi  + B\Theta^2/2 
+ O(\Theta^3) \mbox{ } , \mbox{or}   
\eeq
\beq
2\{\widetilde{\fU} + \widetilde{\fE}\} = 
2\fE - K_2 - L{\cal R}\mbox{cos}\Phi - \{4\fE + K_1 - 3K_2\}{\cal R}^2 + O({\cal R}^3) \equiv 
Q_0 - L{\cal R}\mbox{cos}\Phi  - \{Q_2 {\cal R}^2\} + O({\cal R}^3)
\mbox{ } .  
\label{assmall}
\eeq
Thus the leading term is a constant, 
unless $Q_0 = 2\fE - K_2 \mbox{ }(\mbox{ } \propto \overline{\fE} - A - B \mbox{ }) = 0$, 
in which case it is linear in $\Theta$ or ${\cal R}$ and with a cos$\Phi$ factor, 
unless also $L \mbox{ }(\mbox{ } \propto C \mbox{ }) = 0$ (which is also the condition for the `special' case), 
in which case it is quadratic in $\Theta$ or ${\cal R}$, unless $B = 0$ (given previous conditions, 
this is equivalent to $Q_2 = 4\fE + K_1 - 3K_2 = 0$), which means that one is in the $K_2 = 2\fE$ subcase of 
the `very special' case, for which $\fU + \fE$ has no terms at all.

${\cal R}$ large corresponds to the supplementary angle $\Xi \equiv \pi - \Theta$ being small, so 
\beq
\overline{\fU} + \overline{\fE} = \{\overline{\fE} - A - B\} + C\Xi\mbox{cos}\Phi + B\Xi^2/2 
+ O(\Xi^3) 
\mbox{ } , \mbox{ or }
\eeq
\beq
2\{\widetilde{\fU} + \widetilde{\fE}\} = \{2\fE - K_1\}{\cal R}^{-4} - 
L{\cal R}^{-5}\mbox{cos}\Phi - \{4\fE + K_2 - 3K_1\}{\cal R}^{-6} + O({\cal R}^{-7}) \equiv 
{Q_4}{\cal R}^{-4} - L{\cal R}^{-5}\mbox{cos}\Phi - Q_6{\cal R}^{-6} 
+ O({\cal R}^{-7}) \mbox{ } .  
\label{aslar}
\eeq
Thus the leading term goes as a constant in $\Xi$ or as ${\cal R}^{-4}$, 
unless $Q_4 = 2\fE - K_1 \mbox{ }(\mbox{ }\propto \overline{\fE} - A + B \mbox{ }) = 0$, 
in which case it goes linearly in $\Xi$ or as ${\cal R}^{-5}$ in each case also with a cos$\Phi$ factor, 
unless also $L \mbox{ }(\mbox{ } \propto C \mbox{ }) = 0$ (`special' case), 
in which case it goes quadratically in $\Xi$ or as ${\cal R}^{-6}$, unless $B = 0$ 
(given previous conditions, this is equivalent to $Q_6 = 4\fE + K_2 - 3K_1 = 0$), 
which means that one is in the $K_1 = 2\fE$ subcase of the `very special' case, for which $\fE + \fU$ 
has no terms at all. 
Note that $B = 0$ implies $K_1 = K_2$, so this very special subcase indeed coincides with the previous 
paragraph's.   
Finally note that the large and small asymptotics are dual to each other (the difference of 4 powers is 
accounted for by how the kinetic energy scales under the duality map), so that {\sl one need only 
analyse the parameter space for one of the two regimes and then obtain everything about the other regime 
by simple transcription}.

\subsection{Classical equations of motion}

The Jacobi-type action for this problem is, in spherical coordinates and using the barred 
banal conformal representation, 

\noindent
\beq
\fS  = \int\d\lambda\sqrt{\frac{\dot{\Theta^2} + \mbox{sin}^2\Theta\dot{\Phi}^2}{2}
\left\{ 
\frac{\fE}{4} - A - B\mbox{cos}\Theta - C\mbox{sin}\Theta\mbox{cos}\Phi 
\right\}} \mbox{ } . 
\eeq
Then the Euler--Lagrange equations are
\beq
\Theta^{\overline{**}} - \mbox{sin}\Theta\mbox{cos}\Theta \{\Phi^{\overline{*}}\}^2 = 
B\mbox{sin}\Theta - C\mbox{cos}\Theta\mbox{cos}\Phi \mbox{ } ,
\label{Tera}
\eeq
\beq
\{\mbox{sin}^2\Theta\Phi^{\overline{*}}\}^{\overline{*}} = C\mbox{sin}\Theta\mbox{sin}\Phi \mbox{ } .  
\label{Dactyl}
\eeq
On account of the action being independent of $\overline{t}$, one of these can be replaced by the energy 
integral 
\beq
\frac{\{\Theta^{\overline{*}}\}^2 + \mbox{sin}^2\Theta\{\Phi^{\overline{*}}\}^2}{2} + 
A + B\mbox{cos}\Theta + C\mbox{sin}\Theta\mbox{cos}\Phi = \overline{\fE} \mbox{ } .  
\label{EnIn}
\eeq
A dynamical systems and phase space analysis of these equations and their interpretation in terms of 
$\mI_1$, $\mI_2$ and $\Phi$ will be presented elsewhere \cite{Dyn}.  
Also note that the notions of $\fV_{\mbox{\scriptsize eff}}$ and $\fV_{\mbox{\scriptsize orb}}$ are 
inapplicable in the nonseparable case.

\subsection{`Special case'}

For $C = 0$, 
(\ref{EnIn}) reads 
\beq
\frac{1}{2}
\left\{
\{\Theta^{\overline{*}}\}^2 + \frac{{\cal J}^2}{\mbox{sin}^2\Theta}
\right\} 
+ A + B\mbox{cos}\Theta = \overline{\fE}
\eeq

\noindent
which can be integrated [using (\ref{CF}) to eliminate $t^{\se\sm}$ in favour of $\Phi$]:
\beq
\Phi - \Phi_0 = \int {{\cal J}\d\Theta}/
{\mbox{sin}\Theta\sqrt{2\{\overline{\fE} - A - B\mbox{cos}\Theta\}\mbox{sin}^2\Theta - {\cal J}^2}} 
\mbox{ } .   
\eeq 

\mbox{ }

For the very special case $B = C = 0$, this reduces to the well-known integral for the computation of 
geodesics for the sphere -- c.f. \cite{TriCl}, which exact solution I now cast below in terms of 
${\cal R}$, $\Phi$ and then in terms of the `original' $\underline{\iota}_1$, $\underline{\iota}_2$ 
variables.  
I've also solved the $B \neq 0$ case exactly by e.g. tan$\frac{\Theta}{2} = {\cal R} \equiv \sqrt{x - 1}$, 
obtaining then via Maple \cite{Maple} a composition of polynomials, roots and elliptic functions in $x$
%
%
that I consider to be too complicated to present here.

\cite{TriCl} sketches $\widetilde{\fV}_{\se\sf\sf}$ and $\widetilde{\fU}_{\so\sr\sb}$ for 
the single harmonic oscillator and their large and small asymptotic behaviours.  
${\cal J} = 0$ is rather simpler: 1-d motion i.e. the orbits are straight lines.  
Assume ${\cal J} \neq 0$ from now on in looking for orbits with less trivial shapes. 
The universal large ${\cal R}$ asymptotic solution's $\widetilde{\fV}_{\se\sf\sf}$ 
exhibits a finite potential barrier and the orbits are bounded from above. 
The small ${\cal R}$ asymptotic solution for a fixed $\fE > 0$ 
harmonic oscillator problem is the usual radial/isotropic harmonic oscillator problem.  
Here, $\widetilde{\fV}_{\se\sf\sf}$ is an infinite well formed by the harmonic oscillator's parabolic potential on the 
outside and the infinite centrifugal barrier on the inside, and the orbits are bounded from below.  
For a fixed $\fE > 0$ harmonic oscillator scalefree triangleland problem, in contrast with the usual radial 
harmonic oscillator problem, the potential 
tends to 0 rather than $+\infty$ as ${\cal R} \longrightarrow \infty$, and the orbits are bounded from 
both above and below.

\subsection{Exact solution for the `very special case'}

This effectively constant-$\fV$ case is equivalent to the geodesic problem on the sphere.  
Thus it is solved in $\Theta, \Phi$ variables by great circles (first equality)
\beq
F\mbox{cos}(\Phi - \Phi_0) = 2\mbox{cot}\Theta  =  \{1 - {\cal R}^2\}/{\cal R} = 
\{\iota_2^2 - \iota_1^2\}/\iota_1\iota_2  \mbox{ }   
\label{squik}
\eeq  
(following through with variable transformations so as to cast it in terms of straightforward relational 
variables, and using $\left.{\cal J}F = 2\sqrt{2\{\overline{\fE} - A\} -  {\cal J}^2} = 
\sqrt{2\fE - K_2 - 4{\cal J}^2} = \sqrt{Q_0 - 4{\cal J}^2}\right)$.  
The ${\cal R}$ form above can be rearranged to the equation of a generally off-centre circle [Fig 5 a)].

While, in terms of the original quantities of the problem, 
\beq
F^2\{  \underline{\iota}_1 \cdot \underline{\iota}_2  \}^2 + 
||\underline{\iota}_1||^4 + ||\underline{\iota}_2||^4      + 
2F\{  ||\underline{\iota}_1||^2 - ||\underline{\iota}_2||^2  \}
\underline{\iota}_1 \cdot \underline{\iota}_2 \mbox{cos}\Phi_0 = 
\{2 + F^2\mbox{sin}^2\Phi_0\}||\underline{\iota_1}||^2||\underline{\iota_2}||^2 \mbox{ } .  
\eeq
Note 1) this solution is homogeneous in $||\underline{\iota}_1||$, $||\underline{\iota}_2||$ and 
$\sqrt{\underline{\iota}_1\cdot\underline{\iota}_2}$ (in fact it is a fourth-order homogeneous 
polynomial in these quantities).  
These quantities are all invariant under rotation, so the zero total angular momentum constraint is manifestly 
satisfied, while the zero dilational momentum constraint is manifestly satisfied due to the homogeneity. 
[`Homogeneous equation' is another way of expressing `equation in terms of ratios alone'.]     

\noindent 2) It has simpler subcases: $\mbox{sin}\Phi_0 = 0$ gives the merely second-order 
\beq
F\underline{\iota}_1 \cdot \underline{\iota}_2 + 
||\underline{\iota}_1||^2 = ||\underline{\iota}_2||^2 \mbox{ } , 
\eeq
which I sketch in Fig 5, while cos$\Phi_0 = 0$ gives 
\beq
F^2\{\underline{\iota_1} \cdot \underline{\iota}_2\}^2 + 
||\underline{\iota_1}||^4 + ||\underline{\iota}_2||^4  = 
\{2 + F^2\}||\underline{\iota}_1||^2||\underline{\iota}_2||^2 \mbox{ } .
\eeq 

\noindent 3) The solution is invariant under $\underline{\iota}_1 \longleftrightarrow \underline{\iota}_2$
provided that also $\Phi_0$ is shifted by a multiple of $\pi$.  

\mbox{ } 

\noindent{\footnotesize[{\bf Figure 5} a) On the sphere the solutions are great circles, while their 
projections onto the $({\cal R}, \Phi)$ plane are circles.  
I also provide a sketch of how the projection works;  
the equator maps to the circle ${\cal R} = 1$ and the `Greenwich meridian' maps to the vertical line.  

\noindent b) The barycentric partial moments of inertia as functions of $\Phi$ are then [by  
(\ref{squik}), that their sum is a constant $\mI$, and that this sketch is readily rotatable from 
$\Phi_0 = 0$ to $\Phi_0$ arbitrary and using \cite{Maple}] $\mI_1, \mI_2 = \frac{\sI}{2}\left\{ 1 \pm 
{F\mbox{cos}\Phi}/{\sqrt{4 + F^2\mbox{cos}^2\Phi}}\right\}$.  
This is additionally, as expected, readily rescaleable for any $\mI \neq 0$, so I take $\mI = 1$.  
For diagrams of this kind in this paper, I use the convention of black for $\mI_1$ and dashed for $\mI_2$.  
Then as $F \longrightarrow 0$, one gets twice the circle of radius 0.5 (corresponding to the 
${\cal R} = 1$ solution); the first picture, for $F = 0.1$, is well on its way toward that behaviour.  
As $F$ grows, there motion has two `almost-halves' in which one of $\mI_1$, $\mI_2$ completely 
dominates over the other, separated by two thin wedges in which $\mI_1$, $\mI_2$ cross over.  
This amounts to $\mI_1, \mI_2$ undergoing periodic oscillations.
Note that they always cross over at 1/2, corresponding to all the other circles in Fig 5a) cutting 
the ${\cal R} = 1$ circle.  
Also note the complete symmetry between the status of $\mI_1$ and $\mI_2$ in the opposite direction 
($\Phi \longrightarrow \Phi + \pi$).    
As $F \longrightarrow \infty$, the two curves touch the origin and, oppositely, the unit circle, 
corresponding to ${\cal R}$ running from 0 to $\infty$, i.e. the vertical line of Fig 5a).  

\noindent c) Re-interpretation in terms of particles.   
I consider only $m_1 = m_2 = m_3$ for simplicity.  
The equator (${\cal R} = 1$ circle) $F = 0$ case corresponds to going through the configurations 
sketched, which can be summarized by the picture underneath particles 2, 3 fixed and the circle 
representing the range of positions that particle 1 takes.  
I also provide the analogous summarizing sketch for the slightly inclined great circle of Fig 5 a) 
(this is for $\Phi_0 = 0$, in general different values of $\Phi_0$ correspond to different 
particle behaviours).  
The Greenwich meridian corresponds to motions between the double collision of particles 2, 3 and the 
2, 1, 3 collinearity with particle 1 at the centre of mass of particles 2, 3.  

\noindent d) Some useful terminology for the particle configurations.   
Some definitions to aid this are as follows: 
near-isosceles and near-collinear (for $\alpha <$ some $\alpha_0$, constant), 
`Jacobi-flat' ($\mI_1 << \mI_2$),
`Jacobi-regular' ($\mI_1 = \mI_2$, which, when isosceles, is equilateral) and  
`Jacobi-tall' ($\mI_1 >> \mI_2$).]}

\subsection{Small relative scale asymptotic behaviour}

The first approximation for ${\cal R}$ 
is the generic asymptotics ($Q_0 = 2\fE - K_2 \propto \overline{\fE} - A + B \neq 0$, so 
$\tilde{\fU} + \tilde{\fE}$ goes as $Q_0/2$).  
[Note however that not all dynamical orbits enter such a regime -- sometimes the quadrature's integral 
goes 
complex before the small ${\cal R}$ regime is attained (${\cal R}$ large `classically forbidden') when 
$\widetilde{\fU}({\cal R}) {\cal R}^2 - {\cal J}^2 < 0$.] 
Then integrating the ${\cal R}$ quadrature (\ref{quad}) gives the orbits 
\beq
\pm\sqrt{2q_0}\mbox{sec}(\Phi - \bar{\Phi}) = {\cal R} = \mbox{tan$\frac{\Theta}{2}$} = \iota_1/\iota_2 
\mbox{ }  
\label{exactsoln}
\eeq
(following through with variable transformations so as to cast it in terms of straightforward relational 
variables, and using the ${\cal J}$-absorbing constant 
$q_0 = Q_0/2{\cal J}^2 = {\fE - K_2/2}/{\cal J}^2 = {4\{\overline{\fE} - A - B\}}/{\cal J}^2$).  
Note that the ${\cal R}$ form of the orbits are parallel straight lines (vertical for $\Phi_0 = 0, \pi$) 
[Fig 6a)].    
Moreover, these are known not to be very good approximands in that they totally neglect the 
non-constant part of the potential and thus are precisely rectilinear motions.  
Thus at least here (and one may suspect likewise in the next section), it is often necessary to 
use the second approximation in studies. 
The other equalities in (\ref{exactsoln}) convert that result into what form the asymptotic orbits take 
in straightforward relational variables; I sketch these in Figs 6a) and 7a).

While, in terms of the problem's original quantities, 
\beq
||\underline{\iota}_2||^4 + 2q_0\{\underline{\iota_1} \cdot \underline{\iota_2}\}^2 =  
2q_0\{||\underline{\iota_1}||^2 ||\underline{\iota_2}||^2\mbox{sin}^2\Phi_0 + 
2\sqrt{2q_0}||\underline{\iota}_2||^2\{\underline{\iota_1}\cdot\underline{\iota_2}\}^2\}\mbox{cos}{\Phi}_0    
\mbox{ } ,   
\label{81} 
\eeq 
which is again a fourth-order homogeneous polynomial in $||\underline{\iota}_1||$, 
$||\underline{\iota}_2||$ and $\sqrt{\underline{\iota}_1\cdot\underline{\iota}_2}$.  
It admits the simpler cases 1) $\mbox{sin}\Phi_0 = 0$ (which is merely second-order):  
\beq
||\underline{\iota}_2||^2 = \sqrt{2q_0}\underline{\iota}_1 \cdot \underline{\iota}_2 \mbox{ } , 
\label{simp}
\eeq
and 2) $\mbox{cos}\Phi_0 = 0$:
\beq  
||\underline{\iota}_2||^4 =  
2q_0\{||\underline{\iota}_1||^2||\underline{\iota}_2||^2 - 
      \{\underline{\iota}_1 \cdot \underline{\iota}_2\}^2 \} \mbox{ } .  
\eeq


The second approximation for ${\cal R}$ small in which $L = 0$ but the $Q_2{\cal R}^2/2$ term is also 
kept turns out to also often be necessary.    
If $Q_0 = 0$ but $L \neq 0$ one has non-generic asymptotics not directly covered in this SSec, but 
one can use the technique of Sec 5 to reduce this case to one of those covered in the present SSec 
in a new set of variables.  
If $Q_0, L = 0$, one resides within the special case, and the next leading term is $Q_2{\cal R}^2/2$, 
which case is included in my working below; indeed now the ``second" approximation is always necessary.  
If $Q_2 = 4\fE + K_1 - 3K_2 = 16\{\overline{\fE} - A - 2B\} = 0$ also, one is within the very 
special case, and so one does not need any asymptotic calculations as one has the exact solution of 
SSec 4.8.

For the second approximation integrating the ${\cal R}$ quadrature gives the orbits 
\beq
\pm 1/\sqrt{q_0 + \sqrt{q_0^2 - q_2} \mbox{cos}(2\{\Phi - \bar{\Phi}\})} = {\cal R} = \mbox{tan$\frac{\Theta}{2}$} = \iota_1/\iota_2 
\mbox{ }  
\label{Sexactsoln}
\eeq
(following through with variable transformations so as to cast it in terms of straightforward relational 
variables, and using the ${\cal J}$-absorbing constant $q_2 = Q_2/{\cal J}^2$).  
This is straightforwardly rearrangeable into quite a standard form (e.g \cite{Moulton, Whittaker})
\beq
{\cal R}^2 = \frac{1}{q_0 + \sqrt{q_0^2 - q_2}\mbox{cos}(2\{\Phi - \Phi_0\})} \mbox{ } ,
\eeq
the case-by-case analysis of which is provided in Fig 6b).

While, in terms of the problem's original quantities, and using $g$ for $\sqrt{q_0^2 - q_2}$, 
$$
||\underline{\iota}_2||^8 + \{q_0^2 + 
g^2\mbox{cos}^2(2\Phi_0)\}||\underline{\iota}_1||^4||\underline{\iota}_2||^4 
-2q_0||\underline{\iota}_2||^6||\underline{\iota}_1||^2 + 
4g^2\{\underline{\iota_1} \cdot \underline{\iota_2}\}^4 + 
$$
\beq
2g||\underline{\iota}_2||^2\{q_0||\underline{\iota_1}||^2 - ||\underline{\iota_2}||^2\}
\{2\{\underline{\iota}_1\cdot\underline{\iota}_2\}^2 - 
  ||\underline{\iota}_1||^2||\underline{\iota}_2||^2\}\mbox{cos}(2\Phi_0) = 
4g^2\{\underline{\iota_1}\cdot\underline{\iota_2}\}^2||\underline{\iota}_1||^2||\underline{\iota}_2||^2
\mbox{ } ,
\eeq
which is an eighth-order homogeneous polynomial in $||\underline{\iota}_1||$, 
$||\underline{\iota}_2||$ and $\sqrt{\underline{\iota}_1\cdot\underline{\iota}_2}$.  
It admits the simpler cases 1) $\mbox{sin}(2\Phi_0) = 0$ (which is merely fourth-order):  
\beq
||\underline{\iota}_2||^4 = \{q_0 - g\}||\underline{\iota}_1||^2||\underline{\iota}_2||^2 +
2g\{\underline{\iota}_1\cdot\underline{\iota}_2\}^2
\eeq
and 2) $\mbox{cos}(2\Phi_0) = 0$:
\beq  
||\underline{\iota}_2||^8 +  q_0^2||\underline{\iota}_1||^4||\underline{\iota}_2||^4 
- 2q_0||\underline{\iota}_2||^6||\underline{\iota}_1||^2 = 
4g^2
\{\underline{\iota}_1\cdot\underline{\iota}_2\}^2
\{||\underline{\iota}_1||^2||\underline{\iota}_2||^2 
- \{\underline{\iota}_1\cdot\underline{\iota}_2\}^2  \} \mbox{ } .
\eeq
Taking $q_0 = 0$ further simplifies the simple subcases. 
While, taking $g = 0$ collapses the general solution to
\beq
||\underline{\iota}_2|| = 0 \mbox{ (2, 1, 3 collinearity with 1 at the centre of mass of 2, 3) or} 
||\underline{\iota}_2|| = \pm\sqrt{q_0}||\underline{\iota}_1|| \mbox{ (rectilinear motion: $\Phi$ fixed) } .   
\eeq

\mbox{ } 

\noindent{\footnotesize[{\bf Figure 6.} a) The vertical straight lines of the first small approximation. 

\noindent b) $q_0 > 0$, $q_2 > 0$ (which is the usual mathematics of the isotropic harmonic oscillator), gives ellipses 
centred about the origin \cite{PrincipiaI}, up to maximum values on the $q_2 = q_0^2$ parabola on which 
they become circles centred on the origin; beyond this the solution ceases to exist.  
While, $q_0 > 0, q_2 = 0$ recovers pairs of straight lines belonging to a).  
$q_2 < 0$ is the upside-down isotropic harmonic oscillator and gives hyperbolae, which are obtuse for $q_0 > 0$ 
rectangular on $q_0 = 0$ and acute for $q_0 < 0$.  
Finally, $q_0 = q_2 = 0$ is the circle at infinity, while $q_0 < 0$, $q_2 \geq 0$ and $q_0 = 0, q_2 > 0$ 
are classically forbidden, as can be seen from the quadrature having to be real. 

\noindent c) The partition of parameter space into these cases;  it is obvious from (50) 
that the shaded region is classically disallowed.]}

\mbox{ } 

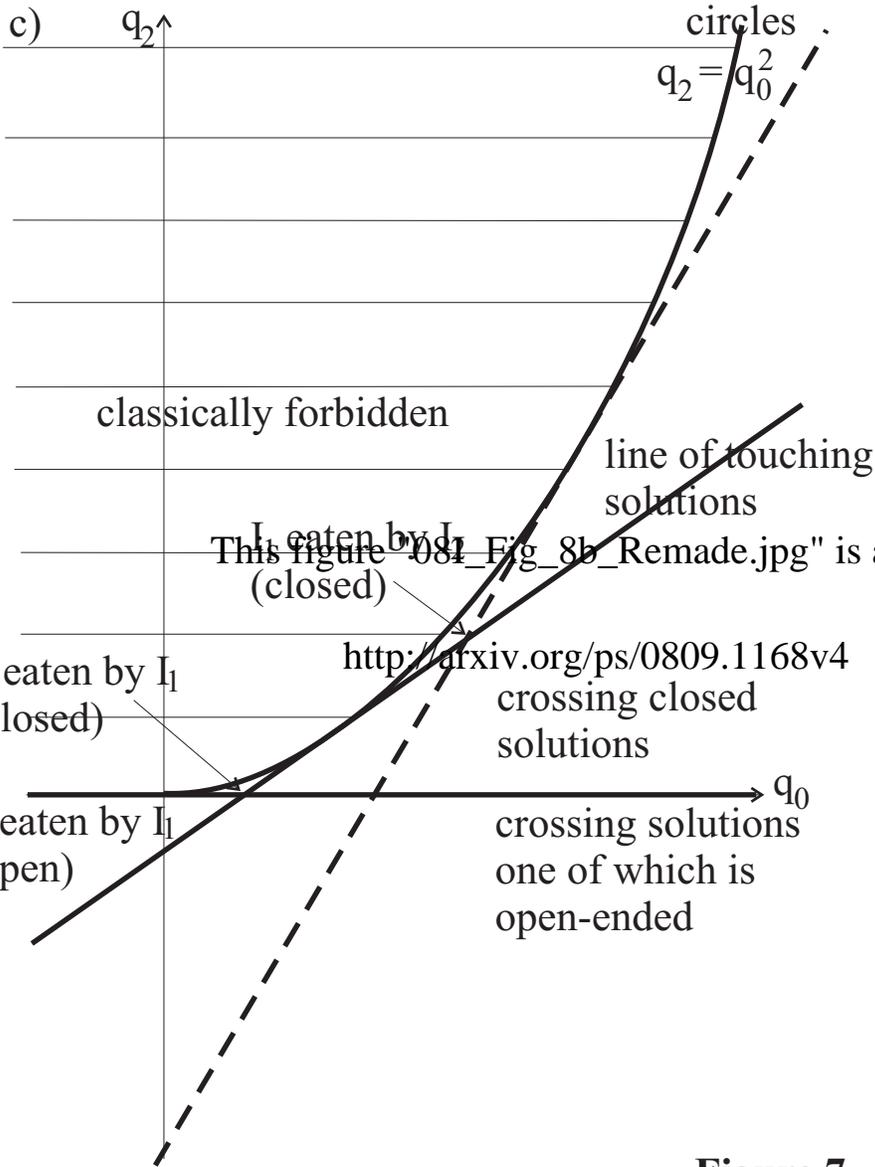
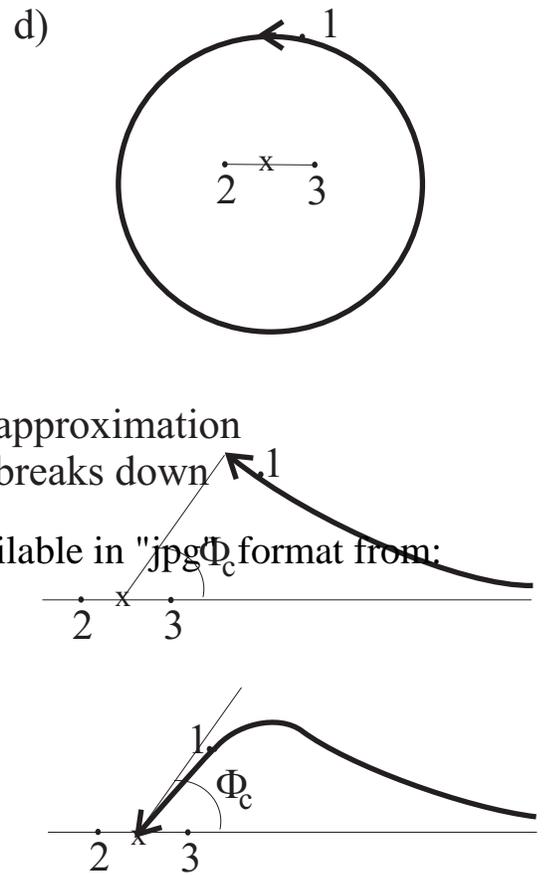
\noindent{\footnotesize[{\bf Figure 7.}  Using a form valid for both first and second small  
approximations, the partial moments of inertia are, for $\Phi_0 = 0$, 

\noindent  
$\mI_1 = {\mI}/\{1 + q_0 + g\mbox{cos}(2\Phi)\}$ and 
$\mI_2 = {\mI\{q_0 + g\mbox{cos}(2\Phi)\}}/\{1 + q_0 + g\mbox{cos}(2\Phi)\}$.   

\noindent a) Then the three types of behaviour of the first large approximation are: `$\mI_2$ is 
surrounded by $\mI_1$', `$\mI_1$ and $\mI_2$ touching', and `$\mI_1$ and 
$\mI_2$ cross-over' (of which I provide two representatives).  
The touching case is the limiting case between the other two cases.  
The only instance in which the small approximation is self-consistent is in the last picture for 
$\Phi$ within wedges around five angular tick-marks wide either side of 0 and of $\pi$.

\noindent b) Next, here are sketches of the nine further behaviours exhibited by the second approximation.  
The first three are limiting behaviours on the $q_2 = q_0^2$ parabola, of which the second is 
itself the limiting behaviour between the other two a single point (corresponding again to 
the ${\cal R} = 1$ solution).
The next three are the unfolded version of `cross-over', the reverse case of `touching' (again a limiting 
behaviour) and the reverse case of `is surrounded'.  
The last four are all cases in which there occurs 
2, 1, 3 collinearity with particle 1 at the centre of mass of particles 2, 3:
`is surrounded', a `touching' limiting case, and two instances of `cross-over'.  
The outwards-lying arrows indicate for which $\Phi$ in each case the small approximation 
is self-consistent.
 
\noindent c) Here I provide a sketch (not to scale) of which regions of the parameter space these 
various cases reside in.
The dashed curve touches the parabola, thus trisecting the classically-allowed region into: 
a region above and to the left of the dashed curve where the small asymptotics is everywhere-valid, 
a region to the right of the dashed curve where it is valid in some wedges, 
and a region below and to the left of the dashed curve where it is nowhere-valid.  

\noindent d) A few examples of reading off what is happening in the particle position picture. 
The third and sixth cases in 7b) correspond to particle 1 describing a closed curve round particles 
2, 3.  
The fifth case describes particle 1 coming in from infinity (= particles 2, 3 colliding) and the 
approximation breaking down for some $\Phi_c$.  
The tenth case describes particle 1 coming in from infinity and reaching the centre of mass of 
particles 2, 3 at some angle $\Phi_c$.]}

 
\subsection{Large relative scale asymptotic behaviour} 

For analogous notions of first and second {\sl large} approximations, now the quadrature in 
${\cal U} = 1/{\cal R}$ takes the same form as the ${\cal R}$-quadrature in the above workings with 
\beq
Q_0 \longrightarrow Q_4 \mbox{ } , \mbox{ } Q_2 \longrightarrow Q_6
\eeq
($q_4$, $q_6$ and $f$ below are then the obvious analogues of $q_0$, $q_2$ and $g$).  
Hence the solutions are dual to those of SSec 4.9's.  
Thus all of SSec 4.9's results apply again under the duality substitutions (and the 
subsequently-induced language changes `small' $\longrightarrow$ `large', and  
``2, 1, 3 collinearity with particle 1 at the centre of mass of particles 2, 3" $\longrightarrow$ 
``collision between particles 2, 3, also interpretable as particle 1 escaping to infinity"). 
In particular, the $\mI_1, \mI_2$ plots of Fig 7b) and which parts of the parameter space the various 
cases hold in and where the now small asymptotics is valid can just be read off the existing figures 
under these substitutions, and then the region in which small asymptotics applied before [Fig 7c)] is 
now that in which large asymptotics now applies.

New ${\cal R}, \Phi$ plots are, however, required (Fig 8)  
Some particular comments are: 1) the first approximation is then 
\beq
\pm\sqrt{2q_4}\mbox{cos}(\Phi - \Phi_0) = {\cal R} = \mbox{tan$\frac{\Theta}{2}$} = \iota_1/\iota_2 \mbox{ } .
\eeq
In the (${\cal R}, \Phi$) plane and for $\Phi_0 = 0$, this takes the form of a family of circles of radius 
$\sqrt{q_4/2}$ and centre $(\sqrt{q_4/2}, 0)$, so that they are all tangent to the vertical axis through 
the origin (\cite{TriCl} and Fig 6a)).  
2) The second approximation gives 
\beq
\pm\sqrt{q_4 + \sqrt{q_4^2 - q_6} \mbox{cos}(2\{\Phi - \bar{\Phi}\})} = {\cal R} = 
\mbox{tan$\frac{\Theta}{2}$} = \iota_1/\iota_2 \mbox{ } .  
\eeq
In the ${\cal R}, \Phi$ plane and for $\Phi_0 = 0$, this take the forms in Fig 8b) in the parameter 
regions delineated by Fig 6c).  

\mbox{ }  

\noindent{\footnotesize[{\bf Figure 8} a) The first large approximation's family of tangent circles in the 
(${\cal R}, \Phi$) plane. 

\noindent b) Additional behaviours of the second large approximation: 
bulging, rectangular and thin tear drops, ellipse-like and peanut-like curves and circles 
centred on the origin.  
The partition of $q_4$, $q_6$ parameter space here is the same as that of $q_0$, $q_2$ 
parameter space in Fig 6c); indeed Fig 6c) is an important clarification of the behaviour 
exhibited by the large approximation.]}  

\mbox{ }  

Note: for general $\widetilde{\fV}_{(\alpha, 0)}$ with constant of proportionality $\Lambda_{\alpha}$, one 
gets exactly the same large-asymptotics analysis as here, with $q_0 = 2\fE - \Lambda_{\alpha}$, 
$q_2 = 4\fE - \{4 + \alpha\}\Lambda_{\alpha}$.  
Thus generally this duality to the isotropic harmonic oscillator of the universal large-scale asymptotics of scalefree 
triangleland is a useful and important result for this Machian mechanics, for the classical and quantum 
mechanics of isotropic harmonic oscillators is rather well-studied and thus a ready source of classical and quantum 
methods, results, and insights.

\subsection{Investigation of the intermediate-${\cal R}$ region}  

Note that only two of $Q_0$, $Q_2$, $Q_4$ and $Q_6$ are independent. 
Thus prescribing a particular small asymptotics entails prescribing a particular large asymptotics too.  
There are however no compatibility restrictions: each small asymptotic behaviour is capable of 
connecting to each large asymptotic behaviour.

Also note from the twin circles, tear drops and peanuts that more than one region in which large 
asymptotics applies is possible.  
Then numerical integration using Maple's \cite{Maple} rkf45 solver reveals that precession can occur 
in the intermediate-${\cal R}$ region, so that orbits can have arbitrarily many large (and small) 
asymptotic regions.  
However, the spiraling reported in \cite{TriCl} does not match this paper's asymptotic 
solutions in all relevant places.  
I account for this as follows.  
1) I observe that when the bona fide energy constant E $\neq 0$, solutions do include the circle centred 
on the origin and spirals to/from that \cite{Moulton} which resemble the previously-reported spiraling.  
2) However the scalefree triangleland problem in hand has E = 0 and so possesses no such solutions.
3) The previously-reported spiraling is due to \cite{TriCl}'s code not implementing the E = 0 
restriction.  
Thus the observation of spiraling with that code is unsurprising, and wrong.  
The code I used for the present paper specifically builds its ${\cal R}$-derivative `initial datum' by 
solving the E = 0 energy equation for the input ${\cal R}$ initial datum (and checks that the evolution 
does not take one away from E = 0), thus amending the error.

\section{Normal coordinates for scalefree triangleland with multi-harmonic oscillator type potentials}

\subsection{A rotation sending the general case to the special case in new coordinates}

One can avoid having a $C$-term by performing a rotation or normal coordinates construction, which 
preserves the form of the kinetic term while sending the potential to a $C$-free form in the new 
coordinates.  
One can get to these by inserting such a rotation between Step 1 and Step 2 in Sec 3.2, or by 
rotating coordinates at the level of the equations of motion.  
I denote the new coordinates with N-subscripts (N for normal).   
Due to the general case in N-subscripted coordinates taking the same form as the special case in the 
original coordinates, we can uplift Sec 4 to an exact solution of general multiple harmonic oscillator case by inserting 
N-subcripts and, at the end of the calculation, rotating back from normal Jacobi coordinates to a more 
complicated form in terms of the original Jacobi coordinates.

What is the requisite rotation angle? 
From the matrix equation 
$\underline{\underline{R}}  (\mbox{through angle } \alpha \mbox{ about } y \mbox{ axis} )
\underline{n} = \underline{n}_{\sN}$,
\beq
\mbox{\fontsize{1.1cm}{1.1cm}\selectfont (}
\stackrel{    \stackrel{  \mbox{cos$\alpha$} \mbox{ } \mbox{0} \mbox{ } -\mbox{sin$\alpha$} }
                       {  \mbox{0} \mbox{ }  \mbox{ } \mbox{1} \mbox{ }  \mbox{ } \mbox{0}  }    }
                       {  \mbox{sin$\alpha$} \mbox{ } 0 \mbox{ } \mbox{cos$\alpha$}  }
\mbox{\fontsize{1.1cm}{1.1cm}\selectfont )}
\mbox{\fontsize{1.1cm}{1.1cm}\selectfont (}
\stackrel{\stackrel{\mbox{$C$}}{\mbox{0}}}{\mbox{$B$}}
\mbox{\fontsize{1.1cm}{1.1cm}\selectfont )}
=
\mbox{\fontsize{1.1cm}{1.1cm}\selectfont (}
\stackrel{    \stackrel{  \mbox{0}  }{  \mbox{0}  }    }{    \mbox{$B_{\sN}$}    } 
\mbox{\fontsize{1.1cm}{1.1cm}\selectfont )} \mbox{ } , 
\eeq
where $\underline{n}$, $\underline{n}_{\sN}$ are unit vectors considered to be in unit-radius spherical 
coordinate form about the original and normal-coordinate axes, the requisite rotation is by 
\beq
\alpha = \mbox{arctan}(C/B) \mbox{ } : 
\eeq
this sends $B\mbox{cos}\Theta + C\mbox{sin}\Theta\mbox{cos}\Phi$ to $B_{\sN}\mbox{cos}\Theta_{\sN}$.  
Then 
\beq 
\mbox{sin}\alpha = {C}/{\sqrt{B^2 + C^2}} \mbox{ } , \mbox{ } 
\mbox{cos}\alpha = {B}/{\sqrt{B^2 + C^2}} \mbox{ } , 
B_{\sN} = \sqrt{B^2 + C^2} \mbox{ } ,    
\eeq
\beq
\mbox{cos}\Theta_{\sN} = \{B\mbox{cos}\Theta + C\mbox{sin}\Theta\mbox{cos}\Phi\}/{\sqrt{B^2 + C^2}}
\mbox{ }
\mbox{ and } 
\mbox{ }
\Phi_{\sN} = \mbox{arctan}
\left(
{\sqrt{B^2 + C^2}\mbox{sin}\Theta\mbox{sin}\Phi}
     /\{B\mbox{sin}\Theta\mbox{cos}\Phi - C\mbox{cos}\Theta\}
\right) \mbox{ } .  
\label{Rex}
\eeq

The potential is now
\beq
\{K_1^{\sN}\rho_{1_{\tN}}^2 + K_2^{\sN}\rho_{2_{\tN}}^2  \}/{8\mI} = A_{\sN} + B_{\sN}\mbox{cos}\Theta_{\sN} 
\mbox{ } .
\eeq
It is also useful to note for later use the following coefficient interconversions: 
\beq
A_{\sN} = A 
\mbox{ } , \mbox{ } ,
K_1^{\sN} = 8\{A - \sqrt{B^2 + C^2}\}
\mbox{ } , \mbox{ }
K_2^{\sN} = 8\{A + \sqrt{B^2 + C^2}\} 
\mbox{ } , 
\label{Lex}
\eeq
\beq
K_1^{\sN} = K_1 + K_2 - \sqrt{\{K_1 - K_2\}^2 + L^2}\} 
\mbox{ } , \mbox{ } 
K_2^{\sN} = K_1 + K_2 + \sqrt{\{K_1 - K_2\}^2 + L^2}\} \mbox{ } .  
\eeq

From the spherical perspective, the normal coordinates solution has the same form as the special 
solution in the original coordinates, but now one is to project onto the general tangent plane rather 
than the tangent plane at the North Pole, interpreting now that general stereographic coordinate 
as the ratio of the square roots of the barycentric partial moments of inertia.  
This permits graphical sketches of the qualitative behaviour rather than fairly lengthy analytical 
expressions constructed by passing from ($\Theta_{\sN}, \Phi_{\sN}$) coordinates to ($\Theta, \Phi$) 
coordinates and then on to mechanically significant variables.

\subsection{Examples: preamble.}

Note that the very special case's solution is invariant under this rotation: that has {\sl no} preferred 
axis.  
So one needs to look slightly further to obtain nontrivial examples.  
First I give the simplest nontrivial example of the first small asymptotics from both an analytical and 
a graphical presentation.  
For this paper to be of manageable length I then only provide a graphical perspective for the first 
large asymptotics, and brief comments on further cases.  

\mbox{ } 

\noindent{\footnotesize[{\bf Figure 9} N-large and N-small domains of applicability map to the original 
(${\cal R}$, $\Phi$) plane as indicated.]}

\subsection{First small asymptotics solution for general case}

From (\ref{exactsoln}) with N-subscripts, tan$\frac{\Theta_{\tN}}{2}$ = $\mbox{sin}\Theta/$\{ 1 + cos$\Theta_{\sN}\}$  
(\ref{Rex}, \ref{Lex}) and elementary cancellations, for $\Phi_{\sN}^{0} = 0$ the analytic solution for this takes 
the form 
\beq
B_{\sN} + B\mbox{cos}\Theta + C\mbox{sin}\Theta\mbox{cos}\Phi = 
\sqrt{2q_{\sN}^0}\{B\mbox{sin}\Theta\mbox{cos}\Phi - C\mbox{cos}\Theta\} \mbox{ } , 
\eeq
where $q_{0}^{\sN}$ bears the same relation to $K_2^{\sN}$ as $q_0$ bears to $K_2$.  
Then in terms of (${\cal R}, \Phi$), 
\beq
\left\{
B_{\sN} - B - \sqrt{2q_{\sN}^0}C
\right\}
{\cal R}^2 + 2
\left\{
{C} - \sqrt{2q_{\sN}^0}B
\right\}
{\cal R}\mbox{cos}\Phi + B_{\sN} + B + \sqrt{2q_{\sN}^0}C  = 0 \mbox{ } .
\eeq
Or, in terms of straightforward relational variables $\mI_1, \mI_2, \Phi$,  
\beq
\left\{
B_{\sN} - B - \sqrt{2q_{\sN}^0}   C
\right\}
\mI_1 + 
\left\{
B_{\sN} + B + \sqrt{2q_{\sN}^0C} 
\right\}
\mI_2 +
2\left\{
C - \sqrt{2q_{\sN}^0}B
\right\}
\sqrt{\mI_1\mI_2}\mbox{cos}\Phi = 0 \mbox{ } .  
\eeq
Finally in terms of the original variables for the problem, 
\beq
\left\{
B_{\sN} - B - \sqrt{2q_{\sN}^0}C
\right\}
||\underline{\iota}_1||^2 + 
\left\{
B_{\sN} + B + \sqrt{2q_{\sN}^0C}
\right\}
||\underline{\iota}_2||^2 + 2
\left\{
{C} - \sqrt{2q_{\sN}^0}B
\right\}
\underline{\iota}_1\cdot\underline{\iota}_2 = 0 \mbox{ } .
\eeq
Like for the $C = 0$ case (\ref{simp}), this is a second-order homogeneous polynomial in 
$||\underline{\iota}_1||$, $||\underline{\iota}_2||$ and 
$\sqrt{\underline{\iota}_1\cdot\underline{\iota}_2}$.  

\mbox{ }

\noindent{\footnotesize[{\bf Figure 10} a) Sketch of how the first small approximation's parallel 
straight lines in the (${\cal R}$, $\Theta$) plane project onto the ($\Theta$, $\Phi$) sphere.  

\noindent b) (${\cal R}$, $\Phi$) sketches for $C \neq 0$, either from plotting the analytical function 
or from projecting the sketch on the sphere onto the tangent plane of the appropriately-rotated 
North Pole. 
The family of parallel straight lines for $C = 0$ is now a family of circles 
tangent to a single point. 
Within the (unshaded) region where this solution is valid, one therefore obtains circular arcs.  

\noindent c) Subsequent sketches of $\mI_1$ and $\mI_2$ as functions of $\Phi$ showing some of the 
distortions which occur when a $C$-term is switched on in the `cross-over' case of Fig 7a).  
All subfigures have $A = 1 = B$ and $\fE$ = 9.
The first picture is for $C = 0.1$.  
The second picture is for $C = 0.5$, by which stage $\mI_2$
%
%
encloses the origin.
The third picture is for $C \approx 2.9$, for which $\mI_1$ and $\mI_2$ touch, 
after which $\mI_2$ `is surrounded' by $\mI_1$ (fourth picture).]}

\subsection{First large asymptotics solution for general case}

\noindent{\footnotesize[{\bf Figure 11} a) Sketch of how the first large approximation's family of 
tangent circles in the (${\cal R}$, $\Theta$) plane project onto the ($\Theta$, $\Phi$) sphere.  

\noindent b) (${\cal R}$, $\Phi$) sketches for C $\neq 0$ either from plotting the analytical function 
or from projecting the sketch on the sphere onto the tangent plane of the appropriately-rotated 
North Pole.
One still has arcs that touch at a point, but this point is no longer at the origin but is rather 
shifted away from it along the $\Phi = 0$ axis.  
This does not cause any noteable changes to $\mI_1$ and $\mI_2$ as functions of $\Phi$, 
so I do not provide sketches of these.]}

\subsection{Further examples}

One can go on from here to provide in rotated coordinates general solutions to the second small and 
large asymptotics, and of the general numerical behaviour in the region in between.   
Here are some brief comments. 

\noindent 1) Similarly to the C = 0 approximation having turn-around ellipses rather than having 
to follow straight lines, the second small approximation allows for turn-around behaviour rather than 
having to complete the first small approximation's circles: approximate circular arc, turn-around, 
approximate  circular arc in opposite direction to the original.  

\noindent 2) One can now have now asymmetric bulges where one had symmetric ones before (e.g. in the 
first large approximation or in the peanut case of the second large approximation); indeed one bulge 
can non-generically become infinitely big (like the straight line in the first large approximation) and 
even form another contribution `beyond infinity' which shows up on the other side of the opposite bulge.

\section{Scaled triangleland at the classical level}

 \subsection{Scale--shape coordinates ($\mI, {\cal R}, \Phi$) or ($\mI, \Theta, \Phi$)}

The general scaled triangleland with multi-harmonic oscillator like potential's Jacobi-type action is  
\beq
\fS_{}[\mI, \Theta, \Phi, \dot{\mI}, \dot{\Theta}, \dot{\Phi}] = 
2\int\d\lambda\sqrt{\check{\fT}\{\check{\fE} + \check{\fU}\}} \equiv  
2\int\d\lambda\sqrt{
\frac{  \dot{\mI}^2 + \mI^2\{\dot{\Theta}^2 + \mbox{sin}^2\Theta\dot{\Phi}^2\}  }{  2  }  
\left\{
\frac{    \fE + \fU(\mI, \Theta, \Phi)    }{    4\mI    }        
\right\}                       }   \mbox{ } .  
\eeq
The Euler--Lagrange equations are 
\beq
\mI^{\check{*} \check{*}} - \mI\{\Theta^{\check{*}2} + \mbox{sin}^2\Theta\Phi^{\check{*}2}\} + \frac{1}{4\mI}
\left\{
\frac{\pa\fV}{\pa \mI} + \frac{\fE + \fU}{\mI}
\right\} = 0 \mbox{ } \mbox{ } , \mbox{ } 
\{\mI^2\Theta^{\check{*}}\}^{\check{*}} - \mI^2\Phi^{\check{*}2}\mbox{sin}\Theta\mbox{cos}\Theta + 
\frac{1}{4\mI}\frac{\pa\fV}{\pa\Theta} = 0  \mbox{ } , 
\label{ERPM_THETA_ELE}
\eeq
\beq
\{\mI^2\mbox{sin}^2\Theta\Phi^{\check{*}}\}^{\check{*}} + \frac{1}{4\mI}\frac{\pa\fV}{\pa\Phi} = 0 \mbox{ } ,  
\label{ERPM_PHI_ELE}
\eeq
and there is an accompanying energy integral 
\beq
\{\mI^{\check{*}2} + \mI^2\{\Theta^{\check{*}2} + \mbox{sin}^2\Theta\Phi^{\check{*}2}\}\}/{2} + 
{\fV}/{4\mI} = {\fE}/{4\mI} 
\mbox{ } .  
\eeq
[Above, 
\beq
\check{*} \equiv \sqrt{\frac{\check{\fE} + \check{\fU}}{\check{\fT}}} \mbox{ } \dot{\mbox{}} = 
\frac{1}{4\mI}\sqrt{\frac{\fE + \fU}{\fT}} \mbox{ } \dot{\mbox{}} = \frac{1}{4\mI} * \mbox{ } .] 
\label{checkstardef}
\eeq

In the case in which $\fV$ is independent of $\Phi$, (\ref{ERPM_PHI_ELE}) simplifies to the first 
integral 
\beq
\mI^2\mbox{sin}^2\Theta\Phi^{\check{*}} = J \mbox{ } .
\eeq
See App C for physical interpretations of the relative angular momentum quantity $J$.

Then the other Euler--Lagrange equations and the energy integral take the forms
\beq
\mI^{\check{*}\check{*}} - \mI\Theta^{\check{*}2} - \frac{  J^{2}  }{  \mI^3\mbox{sin}^2\Theta  } + \frac{1}{4\mI}
\left\{
\frac{\pa\fV}{\pa \mI} + \frac{\fE - \fV}{\mI}
\right\} = 0 \mbox{ } \mbox{ } , \mbox{ } \mbox{ }
\{\mI^2\Theta^{\check{*}}\}^{\check{*}} - \frac{  J^2\mbox{cos}\Theta  }{  \mI^2\mbox{sin}^3\Theta  } + 
\frac{1}{4\mI}\frac{\pa\fV}{\pa\Theta} = 0  \mbox{ } , 
\label{ERPM_THETA_ELE2}
\eeq
\beq
\frac{\mI^{\check{*}2}}{2} + \frac{\mI^2\Theta^{\check{*}2}}{2} + \frac{  J^2  }{  2\mI^2\mbox{sin}^2\Theta  } 
+ \frac{\fV}{4\mI} = \frac{\fE}{4\mI} 
\mbox{ } .
\eeq

\subsection{Scaled triangleland with general multi-harmonic oscillator potential}

The usual general multi-harmonic oscillator potential maps to 
\beq
\{{1}/{4\mI}\}\mI
\left\{
\{K_1 + K_2\}/{4} + \{K_2 - K_1\}/{4}\mbox{cos}\Theta + 
\{{L}/{4}\}\mbox{sin}\Theta\mbox{cos}\Phi
\right\} 
= A + B\mbox{cos}\Theta + C\mbox{sin}\Theta\mbox{cos}\Theta \mbox{ } .
\eeq

\subsection{Sketch of the potential}

The sketch of potential is as for the corresponding similarity problem.  
The sketch of $\check{\fV} - \check{\fE}$ is slightly different from that of $\overline{\fV} - 
\overline{\fE}$ for the similarity problem in that one has a further equation in looking for critical points, 
\beq
{\pa{\{\check{\fV} - \check{\fE}\}}}/{\pa \mI} = {\fE}/{4\mI^2} \mbox{ } , 
\eeq
so one also needs $\fE = 0$ in order to have these.  
After that the analysis is as before (including the picking out of a preferred axis except in the 
$B = C = 0$ case), except that the Hessian has an extra row and column of zeros.

\subsection{Classical equations of motion}

The Jacobi-type action for this problem is then 
\beq
\fS_{} = 2\int\d\lambda\sqrt{
\frac{\dot{\mI}^2 + \mI^2\{\dot{\Theta}^2 + \mbox{sin}^2\Theta\dot{\Phi}^2\}}{2}  
\left\{
\frac{\fE}{4\mI} - A - B\mbox{cos}\Theta - C\mbox{sin}\Theta\mbox{cos}\Phi  
\right\}                       } 
\mbox{ } .  
\eeq
The Euler--Lagrange equations are 
\beq
\mI^{\check{*}\check{*}} - \mI\{\Theta^{\check{*}2} + \mbox{sin}^2\Theta\Phi^{\check{*}2}\} + {\overline{\fE}}/{\mI} = 0 \mbox{ } , 
\label{A}
\eeq
\beq
\{\mI^2\Theta^{\check{*}}\}^{\check{*}} - \mI^2\Phi^{\check{*}2}\mbox{sin}\Theta\mbox{cos}\Theta + 
C\mbox{cos}\Theta\mbox{cos}\Phi - B\mbox{sin}\Theta = 0  \mbox{ } , 
\label{B}
\eeq

\noindent
\beq
\{\mI^2\mbox{sin}^2\Theta\Phi^{\check{*}}\}^{\check{*}} - C\mbox{sin}\Theta\mbox{sin}\Phi = 0 \mbox{ } .
\label{C}
\eeq
The energy integral is
\beq
\frac{    \mI^{\check{*}2} + \mI^2\{\Theta^{\check{*}2} + \mbox{sin}^2\Theta\Phi^{\check{*}2}\}    }{    2    } + 
A + B\mbox{cos}\Theta + C\mbox{sin}\Theta\mbox{cos}\Phi = \frac{\overline{\fE}}{\mI} 
\mbox{ } . 
\label{D}
\eeq

For small ${\cal R}$ or ${\Theta}$ and for large ${\cal R}$ or small $\Xi = \pi - \Theta$, Sec 4's 
results for approximations to the potential hold again.

\subsection{Special Case}

In the special case of $C = 0$, (\ref{ERPM_PHI_ELE}) applies and the remaining Euler--Lagrange equations and energy integral 
become
\beq
\mI^{\check{*}\check{*}} - \mI\Theta^{\check{*}2} -  {  J^{2}  }/{  \mI^3\mbox{sin}^2\Theta  }   +  
{\overline{\fE}}/{\mI} = 0 \mbox{ } \mbox{ } , \mbox{ } \mbox{ } 
\{\mI^2\Theta^{\check{*}}\}^{\check{*}} -  {  J^2\mbox{cos}\Theta  }/{  \mI^2\mbox{sin}^3\Theta  }   + 
C\mbox{cos}\Theta\mbox{cos}\Phi - B\mbox{sin}\Theta = 0  \mbox{ } , 
\label{F}
\eeq
\beq
{\mI^{\check{*}2}}/{2} + {\mI^2\Theta^{\check{*}2}}/{2} + 
{  J^2  }/{  2\mI^2\mbox{sin}^2\Theta  }   + 
A + B\mbox{cos}\Theta + C\mbox{sin}\Theta\mbox{cos}\Phi = {\overline{\fE}}/{\mI} 
\mbox{ } . 
\label{G}
\eeq

\subsection{Analogy between very special case and Kepler--Coulomb problem}

The very special Euclidean relational particle mechanics harmonic oscillator banal-conformally maps to the Kepler problem with 
\beq
\mbox{ (radius) } =  r \mbox{ } \longleftrightarrow \mbox{ } \mI \mbox{ (total moment of inertia) } , \mbox{ } 
\eeq
\beq
\mbox{ (test mass) } =  m \mbox{ } \longleftrightarrow \mbox{ } 1 \mbox{ } , \mbox{ } 
\eeq
\beq
\mbox{ (angular momentum) }   =       \mL \mbox{ } \longleftrightarrow \mbox{ } J \mbox{ (relative angular momentum -- see App C) } , \mbox{ } 
\eeq
\beq
\mbox{ (total energy) } = E \mbox{ } \longleftrightarrow \mbox{ } - A = 
\mbox{ -- (sum of mass-weighted Jacobi--Hooke coefficients)/16 }   
\eeq
and
\beq
\mbox{ (Newton's gravitational constant)(massive mass)(test mass) } = GMm \mbox{ } \longleftrightarrow \mbox{ } \overline{\fE} 
\mbox{ (total energy)/4 }  
\eeq
[or to the 1-electron atom Coulomb problem with this last analogy replaced by 
\beq
\mbox{ (nuclear charge)(test charge of electron)/4$\pi$(permettivity of free space) } = 
(Ze)e/4\pi\epsilon_0 \mbox{ } \longleftrightarrow \mbox{ } \overline{\fE}  \mbox{ (total energy)/4 ] } 
\mbox{ } .
\eeq
Also note that the positivity of the Hooke's coefficients translates to the requirement that the 
gravitational or atomic energy be negative, i.e. to bound states.  
While, the positivity of $\fE$ required for classical consistency corresponds to attractive problems 
like the Kepler problem or the atomic problem being picked out, as opposed to repulsive Coulomb problems.

Also, the special case corresponds to the same `background electric field'  that the rotor was subjected to in 
Sec 4, which, moreover, is proportional to cos$\Theta$ which is analogous to cos$\theta$, which is in the 
axial (`$Z$') direction but is {\sl not} the well-known mathematics of the axial (`$z$') direction 
{\sl Stark effect} for the atom, which involves, rather, $r \mbox{cos}\theta$.  
But, nevertheless, the situation in hand is both closely related to the rotor situation in Sec 4 and 
to the mathematics of the atom in parabolic coordinates (see e.g. \cite{LLQM, Hecht}).  
The general case is then the same situation but with the `electric field' pointing in an arbitrary 
direction.  
The idea then is to use the obvious analogue of the scheme in Fig 3 to solve Euclidean relational particle mechanics problems in straightforward 
relational, relative and absolute terms.

\subsection{Exact solution for the very special case}

Now 
\beq
{  \mI^{\check{*}2}  }/{  2  } + {  J^2  }/{  2\mI^2\mbox{sin}^2{\Theta}_0  } + A = {\fE}/{\mI}
\eeq
[usually one would set $\Theta_0 = \pi/2$ without loss of generality, however the present physical 
interpretation has the value of $\Theta_0$ be meaningful, as 
$\Theta_0 = 2\mbox{arctan}(\iota_1/\iota_2)$].  
Thus the solutions are conic sections  
\beq
\mI = {l}/\{1 + \mbox{$e$ cos}(\Phi - \Phi_0)\}
\eeq
where the semi-latus rectum and the eccentricity are given by 
\beq
l = {J^2}/{\overline{\fE}\mbox{sin}^2\Theta_0}  
\mbox{ } , \mbox{ } 
e = \sqrt{1 - {2AJ^2}/{\overline{\fE}^2\mbox{sin}^2\Theta_0}} \mbox{ } .  
\eeq
So, in terms of straightforward relational variables,  
\beq
\mI_1 + \mI_2 = {l}/\{1 + \mbox{$e$ cos}(\Phi - \Phi_0)\} \mbox{ } , 
\eeq
and in terms of the original variables of the problem,
\beq
\{||\underline{\iota}_1||^2 + 
||\underline{\iota}_2||^2\}\{||\underline{\iota}_1||||\underline{\iota}_2|| + e\{
\underline{\iota}_1\cdot \underline{\iota}_2\mbox{cos}\Phi_0 + 
\sqrt{||\underline{\iota}_1||^2 ||\underline{\iota}_2||^2 - 
      \{\underline{\iota}_1\cdot\underline{\iota}_2\}^2    }\mbox{sin}\Phi_0\}\} = l \mbox{ } .  
\eeq
Note that this is {\sl non}-homogeneous (it rearranges to a non-homogeneous eighth order polynomial);  
this is OK as solutions of {\sl Euclidean} relational particle mechanics do not have to be scale-invariant.

In moment of inertia--relative angle space, for $2A = \{\overline{\fE}\mbox{sin}\Theta_0/J\}^2$ one has 
circles, for $0 < 2A < \{\overline{\fE}\mbox{sin}\Theta_0/J\}^2$ one has ellipses, for $A = 0$ one has 
parabolae (corresponding to the case with no springs).  
The hyperbolic solutions ($A < 0$) are not physically relevant here because this could only 
be attained with negative Hooke's coefficient springs.  
The circle's radius is $\mI = l = \overline{\fE}/2A$ while for the ellipses $\mI$ is bounded to lie 
between $J/\sqrt{2A}\mbox{sin}\Theta_0$ and $\overline{\fE}/2A$.  
The smallest $\mI$ attained in the parabolic case is $J^2/2\overline{\fE}\mbox{sin}^2\Theta_0$.  
The period of motion for the circular and elliptic cases is $\pi\overline{\fE}/\sqrt{2A^3}$.

As regards the individual subsystems, combining the fixed plane equation and the $\mI(\Phi)$ relation,
\beq
\mI_1 = {   l\mbox{sin}^2\mbox{$\frac{\Theta}{2}$}    }/
           \{   1 + \mbox{$e$ cos}(\Phi - \Phi_0)    \} 
\mbox{ } \mbox{ } , \mbox{ } \mbox{ } 
\mI_2 = {   l\mbox{cos}^2\mbox{$\frac{\Theta}{2}$}    }/ 
           \{   1 + \mbox{$e$ cos}(\Phi - \Phi_0)    \} 
\eeq
so each of these behave individually similarly to the total $\mI$.  
In the $\Theta_0 = \pi/2$ plane, they are both equal (and so equal to \mI/2).  
Circle and ellipse cases have $\mI_1$ and $\mI_2$ as closed bounded curves which sit inside the 
curve that $\mI$ traces.  
The parabolic case has $\mI_1$, $\mI_2$ curves to the `inside' of the parabola that $\mI$ traces.

\subsection{Special case solved}

Use $\mI_1, \mI_2, \Phi$ coordinates given by 
\beq
\mI_1 \equiv \{\mI - Z\}/{2} \equiv {\mI}\{1 - \mbox{cos}\Theta\}/2
\mbox{ } \mbox{ } \mbox{ and } \mbox{ } \mbox{ } 
\mI_2 \equiv \{\mI + Z\}/{2} \equiv {\mI}\{1 + \mbox{cos}\Theta\}/2 \mbox{ } ,
\label{Idef}
\eeq
which invert to 
\beq
\mI = \mI_1 + \mI_2 \mbox{ } , \mbox{ } \Theta = \mbox{arccos}
\left(
\{\mI_2 - \mI_1\}/\{\mI_1 + \mI_2\}   
\right)
\eeq
and are mathematically parabolic coordinates scaled by 1/2, which moreover in the present relational 
context have the physical interpretation of partial moments of inertia of the two subsystems.  
Then
\beq
\fS_{} = 2\int \d \lambda \sqrt{\overline{\fT}\{\overline{\fE} + \overline{\fU}\}} = 
2\int\d\lambda\sqrt{\frac{1}{2}
\left\{ 
\frac{\dot{\mI}_1^2}{\mI_1} + \frac{\dot{\mI}_2^2}{\mI_2} + \frac{4\mI_1\mI_2\dot{\Phi}^2}{\mI_1 + \mI_2}
\right\}
\left\{
\frac{\fE}{4} - \frac{K_1\mI_1 + K_2\mI_2}{8} 
\right\}       }
\eeq
for $\overline{\fT}$, $\overline{\fU}$, $\overline{\fE}$ as before.  
Then the $\Phi$-Euler--Lagrange equation is 
\beq
\frac{4\mI_1\mI_2\Phi^{\overline{*}}}{\mI_1 + \mI_2} = J \mbox{ } ,
\eeq
and the energy integral is, subsequently,  
\beq
\frac{\mI_1^{\overline{*2}}}{2\mI_1} + \frac{\mI_2^{\overline{*2}}}{2\mI_2} + \frac{J^2}{8}
\left\{
\frac{1}{\mI_1} + \frac{1}{\mI_2}
\right\} 
+ \frac{K_1\mI_1 + K_2\mI_2}{8} = \frac{\fE}{4} \mbox{ } , 
\eeq
which separates into 
\beq
4\mI_i^{\overline{*}2} + J^2 + K_i\mI_i^2 = 2\fE_i\mI_i 
\eeq
for $\fE_1 + \fE_2 = \fE$.
This is solved by 
\beq
\overline{t} - \overline{t}_0 = \{2/\sqrt{K_i}\}\mbox{arccos}
\left(
\{\mI_iK_i - \fE_i\}/{\sqrt{\fE_i^2 - K_iJ^2}} 
\right)
\eeq
(in agreement with \cite{TriCl}, once differences in convention are taken into account).  
Thus, synchronizing, one part of the equation for the orbits is 
\beq
\sqrt{K_2}\mbox{arccos}
\left(
\{\mI_1K_1 - \fE_1\}/{\sqrt{\fE_1^2 - K_1J^2}} 
\right) = 
\sqrt{K_1}\mbox{arccos}
\left(
\{\mI_2K_2 - \fE_2\}/{\sqrt{\fE_2^2 - K_2J^2}} 
\right) \mbox{ } .  
\eeq
[One can see how the arccosines cancel in the very special case... 
Then $\fE_1 = \fE_2 = \fE/2$ gives  ${\iota_1} = {\iota_2}$ i.e. $\Theta = 
2\,\mbox{arctan} \left(\iota_1/\iota_2\right) = 2$, $\mbox{arctan}(1) = {\pi}/{2}$, 
so are confined to the plane perpendicular to the chosen Z-axis.]  
Then the $\Phi$-Euler--Lagrange equation implies
$$
\Phi - \Phi_0 = J\int\d\overline{t}
\left\{
\frac{1}{\mI_1} + \frac{1}{\mI_2}
\right\} = 
\frac{J}{2}\sum_{i = 1}^{2}\sqrt{K_i}\int\frac{\d\tau_i}{\fF_i\mbox{cos}\tau_i + \fE_i} = 
\sum_{i  = 1}^2 \mbox{arctan}
\left(
\sqrt{\frac{\{\fE_i - \fF_i\}\{\fF_i - \fA_i\}}{\{\fE_i + \fF_i\}\{\fF_i + \fA_i\}}}
\right)
$$  
(for $\tau_i = 2\sqrt{K_i}\{t - t_0\}$, $\fF_i \equiv \sqrt{\fE_i^2 - K_iJ^2}$ and 
$\fA_i = K_i\mI_i - \fE_i$, ), which simplifies to 
\beq
\Phi - \Phi_0 = \sum_{i = 1}^{2}\mbox{arctan}
\left(
\sqrt{\left.\left\{\left\{\sqrt{\fE_i^2 - K_iJ^2} - \fE_i\right\}\mI_i  + J^2\right\}\right/
           \left\{\left\{\sqrt{\fE_i^2 - K_iJ^2} + \fE_i\right\}\mI_i^2  - J^2\right\}   }
\right)
\eeq
in the straightforward relational variables.  
While, in the original variables of the problem, 
\beq
\sqrt{K_2}\mbox{arccos}
\left(
\{||\underline{\iota}_1||^2K_1 - \fE_1\}/{\sqrt{\fE_1^2 - K_1J^2}} 
\right) = 
\sqrt{K_1}\mbox{arccos}
\left(
\{||\underline{\iota}_2||^2K_2 - \fE_2\}/{\sqrt{\fE_2^2 - K_2J^2}} 
\right) \mbox{ } ,  
\eeq
\beq
\mbox{arccos}
\left(
\frac{\underline{\iota}_1\cdot\underline{\iota}_2}{||\underline{\iota}_1||||\underline{\iota}_2||}
\right) = \Phi_0 + \sum_{i = 1}^{2}\mbox{arctan}
\left(
\sqrt{\left.\left\{\left\{\sqrt{\fE_i^2 - K_iJ^2} - \fE_i\right\}||\underline{\iota}_i||^2  + J^2\right\}\right/
            \left\{\left\{\sqrt{\fE_i^2 - K_iJ^2} + \fE_i\right\}||\underline{\iota}_i||^2  - J^2\right\}   }
\right) \mbox{ } .
\eeq

\subsection{The single harmonic oscillator case requires a separate working}

For $K_1 = K_2 = 0$, the trajectories are given by, after some manipulation, 
\beq
\sqrt{    {\fE_2}/{\fE_1}    }\mI_2 = \mI_1 = \mbox{sec}
(\fE_1\{\Phi - \Phi_0\}/\fE)/{\sqrt{2\fE_1}} \mbox{ } ,  
\eeq
which is obviously the expected straight line in the absense of forces.

For $K_2 = 0$, $K_1 \neq 0$, the trajectories are given by, in straightforward relational variables,  
\beq
\{\mI_1K_1  - \fE_1\}/\sqrt{\fE_1^2 - K_1 J^2} = 
\mbox{cos}
\left(
\sqrt{{2K_1}/{\fE_2}}\sqrt{2\fE_2\mI_2 - J^2}
\right)
\label{melenes}
\eeq 
and
\beq
\Phi - \Phi_0 = \mbox{arctan}
\left(
\sqrt{       \left.\left\{\left\{\sqrt{\fE_1^2 - K_1J^2} - \fE_1 \right\}\mI_1  + J^2 \right\}\right/
             \left\{\left\{\sqrt{\fE_1^2 - K_1J^2} + \fE_1 \right\}\mI_1  - J^2  \right\}        }
\right)
+ \mbox{arctan}
\left(
\sqrt{        \frac{2\fE_2\mI_2}{J^2}  - 1   }
\right) \mbox{ } .  
\label{hiperboliques}
\eeq

While, in terms of the original variables of the problem, 
\beq
\{K_1||\underline{\iota}_1||^2  - \fE_1\}/\sqrt{\fE_1^2 - K_1 J^2} = 
\mbox{cos}
\left(
\sqrt{{2K_1}/{\fE_2}}\sqrt{2\fE_2||\underline{\iota}_2||^2 - J^2}
\right) \mbox{ } ,
\label{Omelenes}
\eeq 
$$
\mbox{arccos}
\left(
\frac{\underline{\iota}_1\cdot\underline{\iota}_2}{||\underline{\iota}_1||||\underline{\iota}_2||}
\right) = \Phi_0 + \mbox{arctan}
\left(
\sqrt{        \frac{2\fE_2||\underline{\iota}_2||^2}{J^2}  - 1   }
\right)
$$
\beq
+ \mbox{arctan}
\left(
\sqrt{       \left.\left\{\left\{\sqrt{\fE_1^2 - K_1J^2} - \fE_1 \right\}||\underline{\iota}_1||^2  + J^2 \right\}\right/
             \left\{\left\{\sqrt{\fE_1^2 - K_1J^2} + \fE_1 \right\}||\underline{\iota}_1||^2  - J^2  \right\}        }
\right)
 \mbox{ } .  
\label{Ohiperboliques}
\eeq

\subsection{A brief interpretation of the previous two subsections' examples}

In SSec 6.9's example, $\iota_2$ (or $\mI_2$) makes a good time-standard as the absolute space intuition of it
`moving in a straight line' survives well enough to confer monotonicity.   
It is convenient then to rewrite (\ref{melenes}, \ref{hiperboliques}) as a curve in parametric form 
with $\mI_2$ playing the role of parameter, leading to the plots in Fig 12. 

\mbox{ }  

\noindent {\footnotesize [{\bf Figure 12} a) 3-d plot showing oscillatory behaviour including some 
changes in the size of the relative angle that occur at regular intervals but can involve `sporadic' 
changes in how much the relative angle changes in each interval. (The particular plot given is for the 1 
harmonic oscillator case, with $K_1 = \fE_1 = \fE_2 = 1$ and $J = 0.1$, with $\Phi$ plotted vertically, $\mI_2$ out of 
the page and $\mI_1$ into the page.)
This is because although 1 is `moving in a straight line' in absolute space, the position with respect 
to which its separation is being measured from in relational space (the centre of mass of particles 2, 
3) is then also moving around due to the oscillations of the `spring' between these particles.

\noindent b) Polar plot of $\mI_1$ and $\mI_2$ as functions of $\Phi$ for the first oscillation.  
Further oscillations correspond to similar angular variations at larger radius.     
N.B. that $\mI_1$ and $\mI_2$ are independent in Euclidean relational particle mechanics, as opposed to 
summing to a constant I in similarity relational particle mechanics.  
In the solution exhibited, the particles expand away from triple collision while relative angle varying 
oscillations occur, involving almost-isosceles to almost-collinear changes in shape.]}  

\mbox{ }

In SSec 6.8's example, $\iota_1$ and $\iota_2$ oscillate boundedly, so neither of these 
(or $\mI_1$ or $\mI_2$ is a good clock parameter from the point of view of monotonicity.
There is again some scope for variation in relative angle $\Phi$, including `sporadic' amplitude 
variations.

\subsection{Normal coordinates for scaled triangleland with multi-harmonic oscillator potential}

The working of Sec 5 holds again (using now $x$, $x_{\sN}$ in place of $n$ and $n_{\sN}$ but radii are 
unaffected by rotations and so cancel out giving the same rotation and $\Theta$, $\Phi$ to 
$\Theta_{\sN}$, $\Phi_{\sN}$ coordinate change as before.
For Euclidean relational particle mechanics, one can uplift from the preceding parts of Sec 6 by inserting 
the above extra steps into the triangleland case of Sec 3.1.

The potential is now
\beq
\{K_1^{\sN}\rho_{1_{\tN}}^2 + K_2^{\sN}\rho_{2_{\tN}}^2  \}/{8\mI} = 
A_{\sN} + B_{\sN}\mbox{cos}\Theta_{\sN} 
\mbox{ } .
\eeq
Then the Jacobi-type action for scaled triangleland with general multi-harmonic oscillator potential in shape--scale 
variables is 
\beq
\fS_{} =  2\int\d\lambda \sqrt{
\frac{  \{\dot{\mI}^2 + \mI^2\{\dot{\Theta}_{\sN}^2 + \mbox{sin}^2\Theta_{\sN}\dot{\Phi}_{\sN}^2\}  }{  2  }
\left\{
\frac{\overline{\fE}}{\mI} - A_{\sN} - B_{\sN}\mbox{cos}\Theta_{\sN}
\right\}       } 
\mbox{ } .  
\eeq
The Euler--Lagrange equations and energy integral that follow from this are, respectively and after discovering the 
conserved quantity $J$ and eliminating it (\ref{F}, \ref{G}) with N-subscripts appended.  
The momenta are (\ref{b}) with N's appended in the last two and the Hamiltonian and the energy 
constraint are (\ref{c}) treated likewise.

I have then rotated the above two exact solutions for the special case to obtain solutions to the 
general case, but these are too lengthy to include in this paper.

\section{Conclusion}

\subsection{Results Summary}

Relational particle models are useful as regards the long-standing absolute versus relative motion debate, and 
also due to structural similarities with the geometrodynamical formulation of General Relativity, 
for Problem of Time in Quantum Gravity.    
1- and 2-d relational particle mechanics are tractable due to the simple nature of their configuration space geometries: 
these are, respectively, $\mathbb{S}^k$ and $\mathbb{CP}^k$ for 1- and 2-d similarity relational particle mechanics. 
Additionally, for the 3-particle case of the latter (`scalefree triangleland'), 
$\mathbb{CP}^1 = \mathbb{S}^2$ holds, making this case even more tractable.  
This and its Euclidean relational particle mechanics counterpart (`scaled triangleland') furbish this 
paper's particular examples.

I consider models with general multiple harmonic oscillator type potentials which, as compared with 
the earlier study \cite{TriCl}, include the new feature of relative angular momentum exchange 
between the two constituent subsystems.  
I get there by first considering the `special' subcase (which has no relative angle $\Phi$ dependence in 
its potential and involves {\sl no} relative angular momentum exchange) and its `very special' sub-subcase 
(which has constant potential in one presentation). 
I then identify scalefree triangleland's very special case's mathematics with that of the linear rigid rotor; the 
special case is then analogous to that with a background electric field aligned with its axis.  
The Euclidean relational particle mechanics special and very special cases' mathematics reduces to some of the mathematics that arises 
in the Kepler--Coulomb problem.   
Finally, I use a rotation or normal coordinates construction to cast the general ($\Phi$-dependent, 
relative angular momentum exchanging) case as the special case in the transformed coordinates.  
This has the mathematics corresponding to the analogue background electric field being unaligned with 
the axis.  
In each case I use the standard spherical or planar mathematics that I have been able to cast the 
problem into so as to obtain solutions, and then map back to provide physical interpretation in terms 
of various mechanically-significant quantities that are more intuitively associated with the original 
relational problems: (barycentric partial moment of inertia, $\Phi$) variables, mass-weighted Jacobi 
inter-particle (cluster) coordinates, and by sketches, what the particles themselves are doing.   
In returning to these various levels, the standard spherical, planar and flat space mathematics becomes 
unusual and nontrivial.

Highlights of my results are as follows.   

\noindent 1) The very special multiple harmonic oscillator like potential scalefree triangleland problem is 
straightforwardly soluble on the sphere and retains a manageable form in terms of the underlying 
mechanical variables.

\noindent 2) While the special multiple harmonic oscillator like potential scalefree triangleland problem is also 
classically exactly soluble on the sphere, its form is very complicated, so I just provide more 
manageable large and small asymptotic solutions.

\noindent 3) In the stereographic plane, the small asymptotics solution has the mathematics of the 2-d 
isotropic harmonic oscillator [($k$, constant) $\times$ (radius)$^2$ potential, including the 
upside-down ($k < 0$) and degenerate ($k = 0$) cases], whereby I obtain complete control of it and then 
characterize it in terms of the underlying mechanical variables.  

\noindent 4) The large asymptotics solution maps to the small asymptotics solution again, under 
inversion of the radius, so I also obtain complete control of it and can characterize it in terms 
of the underlying mechanical variables.   
Moreover, this is important beyond the case with harmonic oscillator like potentials, as scalefree triangleland 
exhibits {\sl universal} large-scale behaviour, various cases of which I can now understand through 
their being related by the inversion to the various cases of conic sections that occur for 
$k \mbox{ } \times $(radius)$^2$ potentials.     

\noindent 5) The very special and special multiple harmonic oscillator Euclidean relational particle mechanics problems are also 
classically exactly soluble in the conformally-related flat space in which they repectively take the 
forms of the Kepler--Coulomb problem and a nonstandard composition of parabolic coordinate subproblems 
thereof.

\noindent 6) Then by a rotation or normal coordinates construction, the general relative-angle dependent 
similarity and Euclidean relational particle mechanics multiple HO (type) potential problems are exactly 
soluble because they map to their special cases in the new coordinates.   
However, these solutions are very complicated in terms of the underlying mechanical variables, and as 
such I mostly only provide sketches of some aspects of their behaviour.

\subsection{Further tractable cases: 4-stop and N-stop metrolands}

Because for scalefree 4-stop metroland (relational particle mechanics of 4 particles in 1-d) the $R_j$ 
are different and related to $\Theta$, $\Phi$ in a different way, physically interesting potentials in 
this case generally map to {\sl different} functions of these coordinates for the 4-stop metroland 
interpretation and for the triangleland interpretation.  
Thus one needs new calculations 
%
%
for 4-stop metroland rather than straightforward deduction from this paper's triangleland results.
E.g. the harmonic oscillator type potential maps to 
\beq
\fV = \frac{{K_3}\mbox{cos}^2\Theta + {K_1}\mbox{sin}^2\Theta\mbox{cos}^2\Phi  + 
                                      {K_2}\mbox{sin}^2\Theta\mbox{sin}^2\Phi}{2} = 
\bar{\ttA} + \bar{\ttB}\mbox{cos}(2\Theta) + \bar{\ttC}\mbox{sin}^2\Theta\mbox{cos}(2\Phi)
\eeq
for $\bar{\ttA} = \{K_1 + K_2 + 2K_3\}/8$, $\bar{\ttB} = \{- K_1 - K_2 + 2K_3\}/8$, 
$\bar{\ttC} = \{K_1 - K_2\}/4$. 
Thus this model admits direct analogues of this paper's `special' ($\tt{C} = 0$) and `very special' 
($\tt{B} = \tt{C} = 0$) subcases.

While, generalizing to scalefree N-stop metroland, the multiple harmonic oscillator type potential maps to 
\beq
\fV = \frac{1}{2}\frac{    \sum_{\barp = 1}^{\sn - 1}K_{\barp}{\cal R}_{\barp}^2 + K_{\sn}    }
                      {    \sum_{\barp = 1}^{\sn - 1}{\cal R}_{\barp}^2 + 1                   } = 
\frac{1}{2}\sum_{p = 1}^{\sn}K_{p}n_{p}^2 
\eeq
for $n_{p}$ the unit vector in the Euclidean configuration space $\mathbb{R}^{n}$.  
Scaled N-stop metroland is also of interest \cite{Cones}.

\subsection{Further work and Applications}

This paper's models remain to be studied from the perspective of dynamical systems \cite{Dyn}.
Its principal application at the moment is that many aspects of this work carry over to QM in paper II 
(similarity case) and \cite{08III} (Euclidean case), and on towards the study of many Problem of Time 
strategies (in particular emergent semiclassical time and records theory \cite{EOT, Records, New} but 
also conceivably internal time approaches and histories theory, as well as investigation of various 
Quantum Gravity and Quantum Cosmology applications such as the problem of observables, operator 
ordering, closed-universe effects, finite-universe effects and the study of small inhomogeneities/clumps).  
Some of these applications would benefit from study of further models: with other potentials (for which 
universal large asymptotics results in this paper will be useful), the above 4-stop and N-stop 
metrolands of comparable tractability to this paper, as well as somewhat harder relational particle 
mechanics in 2-d of N $>$ 3 particles (requiring $\mathbb{CP}^{\sN - 2}$ geometry based methods) and 
3-d models that are much harder \cite{Kendall} even for modest values of N.   

\mbox{ }

\noindent{\bf Acknowledgments}

\mbox{ } 

\noindent I thank Dr Julian Barbour, Dr Brendan Foster and Miss Anne Franzen for discussions, 
the two Anonymous Referees for comments, Professor Gary Gibbons for references, 
the organizers of ``Space and Time 100 Years after Minkowski" Conference at Bad Honnef, Germany, 
the Perimeter Institute and Queen Mary's Relativity Group for invitations to speak and hospitality 
and Dr Julian Barbour also for hospitality. 
Peterhouse for funding this work in 2006-08. 
Professors Malcolm MacCallum, Gary Gibbons, Don Page, Reza Tavakol and Jonathan Halliwell,
and Dr's Julian Barbour and Fay Dowker, for support in the furthering of my career.   
My Wife for help with assembling the figures, and my Wife, Alicia, Amelia, Beth, Emma, Emilie, 
Emily, Joshua, Luke, Simeon and Will for keeping my spirits up.    

\mbox{ }

\noindent{\bf\large Appendix A: Mechanics on an in general curved configuration space}

\mbox{ }

\noindent For a general finite theory of the quantities ${\cal Q}_{\sfA}$ with a curved configuration 
space, consider the Jacobi action
\beq
\fS = 2\int\d\lambda\sqrt{\fT\{\fU + \fE\}}
\eeq
where $\fU({\cal Q}_{\sfA})$ is minus the potential energy $\fV({\cal Q}_{\sfA})$, $\fE$ is the total 
energy and $\fT$ is the kinetic energy, 
\beq
\fT = {\cal M}_{\sfA\sfB}\dot{{\cal Q}}^{\sfA}\dot{{\cal Q}}^{\sfB}/2 \mbox{ } ,  
\label{Te}
\eeq
for ${\cal M}_{\sfA\sfB}$ the curved configuration space metric.

This action works as follows.  
For * $\equiv \sqrt{\{\fU + \fE\}/{\fT}} \mbox{ } \dot{\mbox{}}$, the Euler--Lagrange equations are, in geometrical form, 
\beq
{\cal Q}^{\sfA**} + \Gamma^{\sfA}\mbox{}_{\sfB\sfC}{\cal Q}^{\sfB*}{\cal Q}^{\sfC*} = 
- \frac{\pa\fV}{\pa {\cal Q}_{\sfA}} 
\label{123}
\eeq
and there is an energy first integral 
\beq
{\cal M}_{\sfA\sfB}{\cal Q}^{\sfA *}{\cal Q}^{\sfB *}/2 + \fV({\cal Q}^{\sfC}) = \fE \mbox{ } .  
\eeq

\mbox{ }

\noindent{\bf\large Appendix B: Momenta, Hamiltonians and energy constraints}

\mbox{ }

\noindent These are important as regards the passage to Quantum Theory in paper II and \cite{08III}.  

\mbox{ }

\noindent{\bf B.1 General curved configuration space mechanics}

\mbox{ } 

\noindent The conjugate momenta are 
\beq
{\cal P}_{\sfA} = {\cal M}_{\sfA\sfB}{\cal Q}^{\sfB *} \mbox{ } ,   
\eeq
and there is then as a primary constraint the quadratic energy constraint
\beq
\fH \equiv {\cal N}^{\sfA\sfB}{\cal P}_{\sfA}{\cal P}_{\sfB}/2 + \fV = \fE \mbox{ }   
\eeq
for ${\cal N}^{\sfA\sfB}$ the inverse of ${\cal M}_{\sfA\sfB}$ and $\fH$ the Hamiltonian for the system.  
This is propagated by (\ref{123}).   
Such energy constraints that have quadratic but not linear dependence on the momenta are analogous to 
the GR Hamiltonian constraint and carries associated with its form the frozen formalism aspect of the 
Problem of Time \cite{K92, I93}.

\mbox{ } 

\mbox{ }

\mbox{ }

\noindent{\bf B.2 Scaled triangleland in ($\mbox{\boldmath$\iota$}_{\mbox{\scriptsize\bf 1}}$, 
                                  $\mbox{\boldmath$\iota$}_{\mbox{\scriptsize\bf 2}}$, 
                                  $\mbox{\boldmath$\Phi$}$) coordinates}

\mbox{ } 

\noindent The conjugate momenta are 
\beq
P_{\rho}^{i} = \iota_i^* \mbox{ } , \mbox{ } 
P_{\Phi} = {\iota_1^2\iota_2^2}\Phi^*/\{\iota_1^2 + \iota_2^2\} \mbox{ } .  
\eeq
The classical Hamiltonian and energy constraint are then  
\beq
\fH = \frac{1}{2}\sum_{i = 1}^{2}
\left\{
\{P_{\rho}^i\}^2 + \frac{P_{\Phi}^2}{\iota_i^2} 
\right\} 
+ \fV = \fE 
\mbox{ } .  
\eeq
In the $\Phi$-independent case, $P_{\Phi} = J$, constant, so one has furthermore 
\beq
\fH = \frac{1}{2}\sum_{i = 1}^{2}
\left\{
\{P_{\rho}^i\}^2 + \frac{J^2}{\iota_i^2} 
\right\} 
+ \fV = \fE 
\mbox{ } .  
\eeq

\mbox{ }

\noindent{\bf B.3 Scaled triangleland in ($\mI_{\mbox{\scriptsize\bf 1}}$, 
                                  $\mI_{\mbox{\scriptsize\bf 2}}$, 
                                  $\mbox{\boldmath$\Phi$}$) coordinates}

\mbox{ }

\noindent These are useful in the context of a $\Phi$-independent potential energy in the special and 
very special cases, for which the conjugate momenta are 
\beq
P_i = {\mI_i^{\overline{*}}}/{\mI_i} \mbox{ } , \mbox{ } 
P_{\Phi} = {4\mI_1\mI_2\Phi^{\overline{*}}}/{\mI} = J \mbox{ } , \mbox{ constant } .   
\eeq
The Hamiltonian and the energy constraint are then 
\beq
\overline{\fH} \equiv \frac{\mI_1P_1^2}{2} + \frac{\mI_2P_2^2}{2} + 
\frac{J^2}{8}\left\{\frac{1}{\mI_1} + \frac{1}{\mI_2}\right\} + \frac{K_1\mI_1 + K_2\mI_2}{8} = 
\frac{\fE}{4} \mbox{ } .  
\eeq

\mbox{ }

\noindent{\bf B.4 N-paricle d-dimensional preshape space theory and scalefree N-stop metroland}
 
\mbox{ }  

\noindent The conjugate momenta are 
\beq
P_{\barq} = 
\left\{
\prod_{\barp = 1}^{\barq - 1}\mbox{sin}^2\Theta_{\barp}
\right\}
{\Theta}_{\barq}^* \mbox{ } .  
\eeq
The Hamiltonian and the energy constraint are then 
\beq
\fH \equiv \frac{1}{2}\sum_{\barq = 1}^{\sn - 1}
\frac{P_{\barq}^2}{\prod_{\barp = 1}^{\barq - 1}\mbox{sin}^2\Theta_{\barp}} + \fV(\Theta_{\barp}) 
= \fE \mbox{ } .
\eeq

\mbox{ }  

\noindent{\bf B.5 Scalefree N-a-gonland and the exceptional case of triangleland}
 
\mbox{ } 

\noindent The conjugate momenta are 
\beq
{\cal P}_{{\cal R}_{\barp}} = 
\left\{
\frac{\delta_{\barp\barq}}{1 + ||{\cal R}||^2}   - 
\frac{{\cal R}_{\barp}{\cal R}_{\barq}}{\{1 + ||{\cal R}||^2\}^2}
\right\}
{\cal R}_{\barq}^{\widetilde{*}}
\mbox{ } \mbox{ } , \mbox{ } \mbox{ }  
{\cal P}_{{\Theta}_{\tilde{\sp}}} = 
\left\{
\frac{\delta_{\tilde{\sp}\tilde{\sq}}}{1 + ||{\cal R}||^2} - 
\frac{{\cal R}_{\tilde{\sp}}{\cal R}_{\tilde{\sq}}}{\{1 + ||{\cal R}||^2\}^2}
\right\}
{\cal R}_{\tilde{\sp}}{\cal R}_{\tilde{\sq}}\Theta_{\tilde{\sq}}^{\widetilde{*}} \mbox{ } .  
\eeq
The Hamiltonian and the energy constraint are then 
\beq
\fH \equiv \frac{1}{2\{1 + ||{\cal R}||^2\}}
\left\{
\{\delta^{\barp\barq} + {\cal R}^{\barp}{\cal R}^{\barq}\}
{\cal P}_{{\cal R}_{\barp}}{\cal P}_{{\cal R}_{\barp}} + 
\left\{
\frac{\delta^{\tilde{\sp}\tilde{\sq}}}{{\cal R}_{\barp}^2} + 1|^{\tilde{\sp}\tilde{\sq}}
\right\}
{\cal P}_{\Theta_{\tilde{\sp}}}{\cal P}_{\Theta_{\tilde{\sq}}}
\right\}
+ \fV({\cal R}_{\barp}, \Theta_{\tilde{\sp}}) = \fE
\mbox{ } .
\eeq

\noindent For the specific example of scalefree triangleland with harmonic oscillator like potentials in 
stereographic coordinates, the conjugate momenta are 
\beq
p_{\cal R} = {\cal R}^{\widetilde{*}}
\mbox{ } \mbox{ } , \mbox{ } \mbox{ }  
p_{\Theta} = {\cal R}^2\Theta^{\widetilde{*}}
\eeq
and the Hamiltonian and the energy constraint are given by 
\beq
\overline{\fH} \equiv \frac{1}{2}
\left\{
p_{\cal R}^2 + \frac{p_{\Phi}^2}{{\cal R}^2}
\right\} 
+ \frac{K_1{\cal R}^2 + L{\cal R}\mbox{cos}\Phi + K_2}{2\{1 + {\cal R}^2\}^3} = 
\frac{\fE}{\{1 + {\cal R}^2\}^2} \mbox{ } .
\eeq
While, in spherical coordinates, one has
\beq
p_{\Theta} = \Theta^{\overline{*}} \mbox{ } \mbox{ } , \mbox{ } \mbox{ }
p_{\Phi} = \mbox{sin}^2\Theta \Phi^{\overline{*}}     
\label{b}
\eeq
and
\beq
\widetilde{\fH} \equiv \frac{1}{2}
\left\{
p_{\Theta}^2 + \frac{p_{\Phi}^2}{\mbox{sin}^2\Theta}
\right\}
+  A + B\mbox{cos}\Theta + C\mbox{sin}\Theta\mbox{cos}\Phi \mbox{ } = \overline{\fE} \mbox{ } .  
\label{c}
\eeq

\mbox{ }

\noindent{\bf B.6 Scale-shape formulation of scaled triangleland}

\mbox{ }  

\noindent The conjugate momenta are

\noindent
\beq
p_{\sI} = \mI^{\check{*}} \mbox{ } , \mbox{ } 
p_{\Theta} = \mI^2\Theta^{\check{*}} \mbox{ } , \mbox{ }
p_{\Phi} = \mI^2\mbox{sin}^2\Theta\Phi^{\check{*}} \mbox{ } .  
\eeq
The Hamiltonian and the quadratic energy constraint are then 
\beq
\check{\fH} = \frac{p_{\sI}\mbox{}^2}{2} + \frac{p_{\Theta}\mbox{}^2}{2\mI^2} + 
\frac{p_{\Phi}\mbox{}^2}{2\mI^2\mbox{sin}^2\Theta} + \frac{\fV(\mI, \Theta, \Phi)}{4\mI} = 
\check{\fE} \mbox{ } .
\eeq
In the special case, $p_{\Phi} = {\cal J}$. 
The Hamiltonian and the energy constraint are then 
\beq
\check{\fH} = \frac{p_{\sI}\mbox{}^2}{2} + \frac{p_{\Theta}\mbox{}^2}{2\mI^2} + 
\frac{{\cal J}\mbox{}^2}{2\mI^2\mbox{sin}^2\Theta} + \frac{\fV(\mI, \Theta)}{4\mI}  = \check{\fE} 
\mbox{ } .  
\eeq
In particular, for scaled triangleland with multi-harmonic oscillator potential, this is 
\beq
\check{\fH} = \frac{p_{\sI}\mbox{}^2}{2} + \frac{p_{\Theta}\mbox{}^2}{2\mI^2} + 
\frac{p_{\Phi}\mbox{}^2}{2\mI^2\mbox{sin}^2\Theta} + A + B\mbox{cos}\Theta + 
C\mbox{sin}\Theta\mbox{cos}\Phi = \check{\fE} = \frac{\fE}{4\mI} \mbox{ } , 
\label{this}
\eeq
which is close to but not exactly the same as the classical Hamiltonian for an atom in a background 
homogeneous electric field pointing in an arbitrary direction.
And in normal coordinates, the Hamiltonian and the energy constraint are
\beq
\check{\fH} = \frac{p_{\sI}^2}{2} + \frac{p_{\Theta_{\tN}}^2}{2\mI^2} + 
\frac{p_{\Phi_{\tN}}^2}{2\mI^2\mbox{sin}^2\Theta} + A_{\sN} + B_{\sN} \mbox{cos}\Theta 
= \frac{\fE}{4\mI} \mbox{ } ,   
\eeq
which is close to but not exactly the same as the classical Hamiltonian for an atom in a  
background homogeneous electric field pointing in the axial `$z$' direction.  

\mbox{ }

\noindent{\bf \large Appendix C: Physical interpretation of this paper's relative angular momentum quantities}

\mbox{ }

\noindent The relative angular momentum quantity in scaled triangleland in simple relational variables is 
\beq
J = {\mI_1\mI_2}\Phi^*/{\mI} \mbox{ } . 
\eeq 
This is equivalent to the scale--shape spherical polar form 
\beq
J = \mI^2\mbox{sin}^2\Theta \Phi^{\check{*}}
\label{baralda}
\eeq
by (\ref{Idef}) and (\ref{checkstardef}).  
It has the following interpretation. 
\beq
J\mI = \mI_1\mI_2\Phi^* = \mI_1\mI_2\{\theta_2^* - \theta_1^*\} = \mI_1\mL_2 - \mI_2\mL_1 = 
\{\mI_1 + \mI_2\}\mL_2 = - \{\mI_1 + \mI_2\}\mL_1
\eeq
where the fourth equality uses the zero angular momentum constraint, and so, as $\mI_1 + \mI_2 = \mI$, 
\beq
J = \mL_2 = - \mL_1 = \{\mL_2 - \mL_1\}/2 \mbox{ } .
\label{walda}
\eeq
So it is interpretable as the angular momentum of one of the two constituent subsystems, minus the angular momentum of the other or 
half of the difference between the two subsystems' angular momenta, which is a relative angular momentum presentation.

Using spherical coordinates, scalefree triangleland's relative angular momentum quantity is 
\beq
{\cal J} = \mbox{sin}^2\Theta \Phi^{\overline{*}} = \frac{  \mI^2\mbox{sin}^2\Theta  }{  \mI  }
\frac{1}{4\mI}\Phi^* = 
\frac{    \mI^2\mbox{sin}^2\Theta\Phi^{\check{*}}    }{\mI} = \frac{J}{\mI} 
\eeq
by (\ref{overlinestardef}), (\ref{checkstardef}) and (\ref{baralda}).  
Thus, by (\ref{walda}),  
\beq
{\cal J} = J/\mI = \mL_2/\mI = - \mL_1/\mI = \{\mL_2 - \mL_1\}/2\mI \mbox{ } ,
\eeq
but $\mI$ is constant in similarity relational particle mechanics, so this is still, up to a constant 
of proportion, the angular momentum of one of the two constituent subsystems, minus the angular momentum of the other or half of 
the difference between the two subsystems' angular momenta.  
Additionally, it has the dimensions of rate of change of angle, which makes sense since on the sphere 
only angles are meaningful.

In either case, $\Phi$-independence in the potential corresponds to there being no means for angular momentum to be 
exchanged between the subsystem composed of particles 2, 3 and that composed of particle 1.

For scaled triangleland, one can consider the configuration space vector 
\beq
\underline{\mI} = 
\mbox{\fontsize{1.3cm}{1.3cm}\selectfont (}
\stackrel{       \mbox{$\mI$sin$\Theta$cos$\Phi$}       }
         {       \stackrel{  \mbox{$\mI$sin$\Theta$sin$\Phi$}        }
                          {  \mbox{$\mI$cos$\Theta$} \hspace{0.1in}                 }         }
\mbox{\fontsize{1.3cm}{1.3cm}\selectfont )} = 
\mbox{\fontsize{1.3cm}{1.3cm}\selectfont (}
\stackrel{       \mbox{$2\iota_1\iota_2\mbox{cos}\Phi$}         } 
         {       \stackrel{   \mbox{$2\iota_1\iota_2\mbox{sin}\Phi$}        }     
         {       \mbox{$\iota_2^2 - \iota_1^2$}                               }          }
\mbox{\fontsize{1.3cm}{1.3cm}\selectfont )} \mbox{ } .  
\eeq
Then from this and its conjugate momentum $\underline{\mP}$, the vector 
\beq
\underline{\mJ} = \underline{\mI} \cr \underline{\mP} 
\eeq
can be formed, which is conserved in the very special case; $J$ is the axial `$Z$' component of this, 
$\mJ_Z$, while the new components are
\beq
\mJ_X 
= \frac{2}{\iota_1^2 + \iota_2^2}  
\{ \{{\iota_1^2 + \iota_2^2}\}\{ \iota_1\iota_2^* - \iota_2\iota_1^* \}\mbox{sin}\Phi + 
   \{\iota_1^2 - \iota_2^2\}\iota_1\iota_2\mbox{cos}\Phi \Phi^*      \}
\eeq
and
\beq
\mJ_Y 
= \frac{2}{\iota_1^2 + \iota_2^2}  
\{ \{{\iota_1^2 + \iota_2^2}\}\{ \iota_1\iota_2^* - \iota_2\iota_1^* \}\mbox{cos}\Phi + 
   \{\iota_1^2 - \iota_2^2\}\iota_1\iota_2\mbox{sin}\Phi \Phi^*      \} \mbox{ } .
\eeq
Additionally, the configuration space Laplace--Runge--Lenz type vector 
\beq
\underline{\mQ} = \underline{\mP} \cr \underline{\mJ} - \overline{\fE}\underline{\mI}/\mI =
\eeq
\beq
\left\{ 
4\{    \{\iota_1^2 + \iota_2^2\}  \{\iota_1^{*2} + \iota_2^{*2}\}   +  \iota_1^2\iota_2^2\Phi^{*2}   \} 
- \frac{\overline{\fE}}{\iota_1^2 + \iota_2^2}  
\right\}
\mbox{\fontsize{1.3cm}{1.3cm}\selectfont (}
\stackrel{         \mbox{$2\iota_1\iota_2\mbox{cos}\Phi$}         } 
         {         \stackrel{  \mbox{$2\iota_1\iota_2\mbox{sin}\Phi$}        } 
         {         \mbox{$\iota_2^2 - \iota_1^2$}                       }          }
\mbox{\fontsize{1.3cm}{1.3cm}\selectfont )}
- 4 \{ \iota_1^2 + \iota_2^2  \}\{ \iota_1\iota_1^* + \iota_2\iota_2^* \}
\mbox{\fontsize{1.3cm}{1.3cm}\selectfont (}
\stackrel{          \mbox{$\{\iota_1^*\iota_2 + \iota_2^*\iota_1\}\mbox{cos}\Phi - 
                    \iota_1\iota_2\mbox{sin}\Phi \Phi^*$}                                  } 
         {          \stackrel{\mbox{$\{\iota_1^*\iota_2 + \iota_2^*\iota_1\}\mbox{sin}\Phi + 
                               \iota_1\iota_2\mbox{cos}\Phi\Phi^*$}                        }
         {          \mbox{$\iota_2\iota_2^* - \iota_1\iota_1^*$}                           }         }
\mbox{\fontsize{1.3cm}{1.3cm}\selectfont )} 
\eeq
which is also conserved.  
However, this only furbishes one further independent conserved quantity due to the interdependences 
\beq
\underline{\mJ}\cdot\underline{\mQ} = 0 \mbox{ } \mbox{ and } \mbox{ } 
\mQ^2 = \overline{\fE}^2 - 2A\mJ^4/\overline{\fE}^2 \mbox{ } .  
\eeq

The very special multiple harmonic oscillator like potential case of scalefree triangleland also has not just a conserved quantity ${\cal J}$ but a conserved vector 
$\underline{\cal J}$ of which ${\cal J}$ is the axial `$Z$' component (so that 
$\underline{\cal J} = \underline{\mJ}/\mI$).
$\Phi$--independent scalefree triangleland can still have yet more conserved quantities, but these are 
of a rather more complicated nature along the lines described in e.g. \cite{+LRL, Goldstein}.




\begin{thebibliography}{99}
\bibitem{Maple}              This was done using Maple 10 to 12.
\bibitem{PrincipiaI}         This result was first given in its original context in Chapter 10 of Book I of \cite{Principia}.
\end{thebibliography}
\end{document}